\documentclass[12pt]{article}
\usepackage{amsmath,amsfonts,amssymb}
\usepackage[nosort]{cite}
\usepackage{hyperref}
\usepackage{graphicx}
\usepackage{color}
\usepackage{epsfig}
\usepackage{psfrag}	
\usepackage{tikz}
\usetikzlibrary{snakes}
\usepackage{slashed}
\usepackage{ulem}
\usepackage[font=small,labelfont=bf,width=1\textwidth]{caption}
\usepackage{comment}
\usepackage[all]{xy}
\usepackage{multirow}
\usepackage{float}

\usepackage{color}

\newcommand \be{\begin{eqnarray}}
\newcommand \ee{\end{eqnarray}}

\numberwithin{equation}{section}

\DeclareMathOperator{\Tr}{Tr}

\DeclareMathOperator{\sTr}{sTr}

\DeclareMathOperator{\diag}{diag}

\def\bC {\mathbb{C}}

\newcommand{\bea}{\begin{eqnarray}}
\newcommand{\eea}{\end{eqnarray}}
\newcommand{\beq}{\begin{equation}}
\newcommand{\eeq}{\end{equation}}
\newcommand{\bal}{\begin{equation}\begin{aligned}}
\newcommand{\eal}{\end{aligned} \end{equation}}

\newcommand{\address}[1]{\vbox{\center\em#1}}
\renewcommand{\title}[1]{\vbox{\center\huge{#1}}\vspace{5mm}}

\newcommand{\cA}{{\mathcal A}}

\newcommand{\cD}{{\mathcal D}}

\newcommand{\cG}{{\mathcal G}}

\newcommand{\cL}{{\mathcal L}}

\newcommand{\cN}{{\mathcal N}}
\newcommand{\cP}{{\mathcal P}}
\newcommand{\cQ}{{\mathcal Q}}

\begin{document}

\begin{titlepage}
\begin{center}

\vspace*{20mm}

\title{Notes on hyperloops in $\cN=4$ Chern-Simons-matter theories}

\vspace{10mm}

\renewcommand{\thefootnote}{$\alph{footnote}$}

Nadav Drukker,%
\textsuperscript{1,}%
\footnote{\href{mailto:nadav.drukker@gmail.com}
{\tt nadav.drukker@gmail.com}}
Marcia Tenser%
\textsuperscript{2,}%
\footnote{\href{mailto:marciatenser@gmail.com}
{\tt marciatenser@gmail.com}}
and
Diego Trancanelli%
\textsuperscript{2,3}%
\footnote{\href{mailto:dtrancan@gmail.com}
{\tt dtrancan@gmail.com}}

\vskip 3mm
\address{
\textsuperscript{1}%
Department of Mathematics, King's College London,
\\
The Strand, WC2R 2LS London, United-Kingdom}

\address{
\textsuperscript{2}%
Institute of Physics, University of S\~ao Paulo,
\\
05314-970 S\~ao Paulo, Brazil
}

\address{
\textsuperscript{3}%
 Dipartimento di Scienze Fisiche, Informatiche e Matematiche, \\
Universit\`a di Modena e Reggio Emilia, via Campi 213/A, 41125 Modena, Italy \\ \& \\
INFN Sezione di Bologna, via Irnerio 46, 40126 Bologna, Italy}

\renewcommand{\thefootnote}{\arabic{footnote}}
\setcounter{footnote}{0}

\end{center}

\vspace{8mm}
\abstract{
\normalsize{
\noindent
We present new circular Wilson loops in three-dimensional $\cN=4$ quiver Chern-Simons-matter theory on $S^3$. At any given node of the quiver, a two-parameter family of operators can be obtained by opportunely deforming the 1/4 BPS Gaiotto-Yin loop. Including then adjacent nodes, the coupling to the bifundamental matter fields allows to enlarge this family and to construct loop operators based on superconnections. We discuss their classification, which depends on both discrete data and continuous parameters subject to an identification. The resulting moduli spaces are conical manifolds, similar to the conifold of the 1/6 BPS loops of the ABJ(M) theory. }}
\vfill

\end{titlepage}
\tableofcontents


\section{Introduction}

In hindsight, the word ``profusion'' in the title of \cite{Cooke:2015ila} is simultaneously misguided and prophetic: misguided because the profusion referred to in that paper was in fact a simple finite degeneracy; prophetic because three-dimensional conformal field theories have since been found to enjoy an intricate moduli space of line operators preserving varying numbers of supercharges. The finite degeneracy of the $1/2$ BPS loops (also partially recognized in \cite{Ouyang:2015qma}) was soon realized holographically in \cite{Lietti:2017gtc}. However, the degeneracy of less supersymmetric loops uncovered since  (see, for example, \cite{Ouyang:2015iza,Ouyang:2015bmy,Mauri:2017whf,Mauri:2018fsf}) has blossomed into an independent research program whose full scope is still unclear.

After several years with more and more examples of BPS loops being found, the past year has seen  some initial steps to reorganize the subject. First, the roadmap paper \cite{Drukker:2019bev}  reviewed what was known at the time about BPS Wilson loops in three dimensions, introduced some new  formalism and, for the first time, properly addressed the moduli spaces of the $1/6$ BPS Wilson loops in ABJ(M) theory. Second, the moduli spaces of BPS Wilson loops  in $\cN=2$ theories were studied in \cite{drukker2020bps} and identified with quiver varieties. Most  recently, the symmetries of BPS line operators in diverse dimensions were analyzed in \cite{Agmon:2020pde},  where the naturalness of marginal defect couplings in three dimensions was stressed.

This work aims to continue on that path, focusing on $\cN=4$ Chern-Simons-matter theories \cite{Gaiotto:2008sd,Imamura:2008dt,Hosomichi:2008jd,Hama:2011ea}. Though these theories are more constrained than $\cN=2$ theories, they afford more possibilities of preserved sets of supercharges. So, while in $\cN=2$ theories there are only 1/2 and 1/4 BPS loops, for $\cN=4$ we find here loops preserving 1, 2, 4 and 8 supercharges, thus ranging between $1/16$ BPS and $1/2$ BPS (because of conformality the vacuum has 16 supercharges). Compared to ABJ(M) theory, the allowed set of theories is wider and different loops that are equivalent under the $SO(6)$ R-symmetry of ABJ(M)  may be on disconnected branches of the moduli space in the presence of only $SO(4)$ R-symmetry.

With the proliferation of papers on the topic, let us set the scope of this one explicitly. We aim to study here circular Wilson loops in three-dimensional $\cN=4$ theories that are continuously connected by marginal deformations to the usual $1/4$ BPS Gaiotto-Yin (``bosonic'') Wilson loop \cite{Gaiotto:2007qi}. One type of deformation arising in quiver gauge theories involves couplings to the matter fields in bifundamental representations of the gauge groups. This follows closely previous studies of the  moduli spaces of loops in $\cN=2$ theories and in ABJ(M) \cite{Drukker:2019bev,drukker2020bps}. The other deformation is often called the ``latitude'' deformation and is similar to the four-dimensional Wilson loops in \cite{Drukker:2006ga}. This construction uses that $\cN=4$ theories have triplets of bifundamental bilinears, the  moment-maps that generalize the scalar coupling of the Gaiotto-Yin Wilson loop. 

The usual $1/4$ BPS loop involves couplings to scalar bilinears of both the hyper and twisted hyper fields  of the three-dimensional Chern-Simons-matter theories. These couplings break the $SO(4)\simeq SU(2)_L\times SU(2)_R$ to  $U(1)_L\times U(1)_R$, while the latitude deformation further breaks one of the $U(1)$'s (we choose it to be  $U(1)_R$). This is a continuous deformation with a parameter $\theta$ and for generic values of this parameter the loop preserves $1/8$ of the supercharges. Given this as a starting point, we can deform the bosonic loops by introducing couplings to more bifundamental fields through superconnections and get ``fermionic'' operators, in the spirit of \cite{Drukker:2009hy,Drukker:2019bev}. This produces richer moduli spaces of BPS loops, which also include 1/16 BPS operators. As this construction involves in an intimate way the hypermultiplet fields and their $SU(2)$ R-symmetries, we name these operators {\it hyperloops}.

We choose to define our theories on $S^3$ and to support our loops along great circles of this space. What we call ``latitude'' deformation is then a slight misnomer, as for us this deformation only affects the internal space of moment map couplings and not the geometric contour on which these operators are defined. As a consequence, the two supercharges preserved by the latitude loops are not a subset of the four preserved by the Gaiotto-Yin loop, but this can be resolved by a conformal transformation mapping the original circle at the equator to an actual latitude of the $S^3$. The alternative formulation would follow the four-dimensional construction in \cite{Drukker:2007qr} or the ABJ(M) analog in \cite{Cardinali:2012ru} with the operators defined from the start along latitudes and with the matter field couplings dictated by this choice of geometric contour. In that case, the latitude loops would be a subset of loops with arbitrary shapes on $S^2\subset S^3$, all preserving a fixed subset of the supercharges. So, barring the fact that we have made the choice to keep the circle fixed in space, we can view all the loops presented here as continuous deformations of the Gaiotto-Yin loop preserving a subset of the supercharges.

The classification of the hyperloops depends on (\textit{i}) some discrete data and (\textit{ii}) some continuous 
parameters. The discrete data is a choice of vector fields that appear in the (super)connection 
(and their multiplicities) and a subset of the matter fields that we allow to couple to the loop. 
This information can be conveniently conveyed by certain quiver diagrams, as we explain below. 
Restricting to only half of the matter fields enhances supersymmetry and gives hyperloops that 
are (at least) $1/8$ BPS, so there are two choices of which half of these fields to include, with each 
option spanning a separate branch of the moduli space. One can also allow a coupling to all the 
fields at the price of preserving less supersymmetry and obtaining 1/16 BPS operators. The 
continuous parameters are the latitude angle $\theta$ mentioned above, an azimuthal angle 
$\varphi_0$ (which we mostly ignore) 
and the continuous couplings to the matter fields. In the simplest case, there are two or four complex parameters per edge in the quiver, depending on how many supercharges one wants to preserve. These parameters are subject to a global gauge symmetry, reducing the moduli space to a cone, similar to the conifold in the case of ABJ(M) theory \cite{Drukker:2019bev}.

These notes are organized as follows. In Section~\ref{sec:notation} we spell out the details of the theory under consideration and set up the notation. In Section~\ref{sec:bos} we write down the $1/4$ BPS Wilson loop coupled to a single node of the quiver and a $1/8$ BPS generalization in terms of the latitude parameter $\theta$ and the azimuthal angle $\varphi_0$. In Sections~\ref{sec:theta=0} and~\ref{sec:thetanot0} we construct the deformations of the $1/4$ and $1/8$ BPS bosonic loops by couplings them to bifundamental fields, and study their moduli spaces. In Section~\ref{sec:MM} we propose a matrix model that may compute these operators. We conclude in Section~\ref{sec:conclusions} with some outlook and provide details about the supersymmetry transformations and other technicalities in two appendices.


\section{The theory and notation}
\label{sec:notation}

\begin{figure}[H]
\centering
\begin{tikzpicture}
\draw[line width=.5mm] (2,2) circle (7mm);
\draw[line width=.5mm] (6,2) circle (7mm);
\draw[line width=.5mm] (10,2) circle (7mm);
\draw[line width=.5mm] (-1,2)--(1.3,2);
\draw[line width=.5mm] (2.7,2)--(5.3,2);
\draw[line width=.5mm] (6.7,2)--(9.3,2);
\draw[line width=.5mm] (10.7,2)--(13,2);
\draw (2,2) node  []  {$A_{I-1}$};
\draw (6,2) node  []  {$A_{I}$};
\draw (10,2) node  []  {$A_{I+1}$};
\draw (2,1) node  []  {$k$};
\draw (6,1) node  []  {$-k$};
\draw (10,1) node  []  {$k$};
\draw (0,2.4) node  []  {$\bar{q}_{I-2\,a}, \ \bar{\psi}^{\dot{a}}_{I-2}$};
\draw (0,1.6) node  []  {$q^a_{I-2}, \ \psi _{I-2\,\dot{a}}$};
\draw (4.05,2.4) node  []  {$\tilde{q}_{I-1\,\dot{a}}, \ \tilde{\psi}^a_{I-1}$};
\draw (4.05,1.6) node  []  {$\bar{\tilde{q}}^{\dot{a}}_{I-1}, \bar{\tilde{\psi}}_{I-1\,a}$};
\draw (8,2.4) node  []  {$\bar{q}_{I\,a}, \ \bar\psi^{\dot{a}}_{I}$};
\draw (8,1.6) node  []  {$q^a_{I},\ \psi_{I\,\dot{a}}$};
\draw (12,2.4) node  []  {$\tilde{q}_{I+1\,\dot{a}},\ \tilde{\psi}^a_{I+1}$};
\draw (12,1.6) node  []  {$\bar{\tilde{q}}^{\dot{a}}_{I+1},\ \bar{\tilde{\psi}}_{I+1\,a}$};
\end{tikzpicture}
\caption{The quiver and field content of the ${\cal N}=4$ theory.}
\label{fig:N=4quiver}
\end{figure}
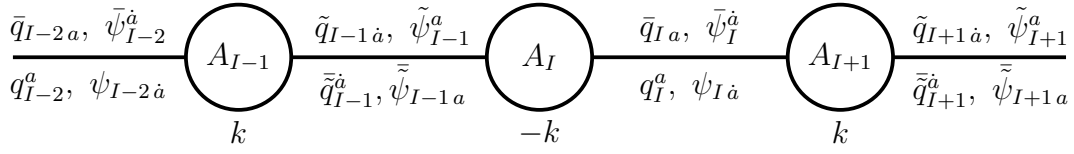

We consider an $\cN=4$ Chern-Simons-matter theory, whose quiver is either circular or linear. 
For the most part we discuss a node labeled by $I$ with gauge field $A_I$, and the adjacent 
nodes with $A_{I\pm1}$. There is a hypermultiplet $(q_I^a,\psi_{I\dot a})$ coupling to $A_I$ 
and $A_{I+1}$ and a twisted hypermultiplet $(\tilde q_{I-1\,\dot a},\tilde\psi_{I-1}^{a})$ coupling 
to $A_I$ and $A_{I-1}$, and so on in an alternated fashion. 
The field content is summarized in the quiver diagram of Figure~\ref{fig:N=4quiver}, where 
the thick solid lines represent the matter fields.

\begin{figure}[H]
\centering
\begin{tikzpicture}
\draw[line width=.5mm] (2,2) circle (7mm);
\draw[line width=.5mm] (6,2) circle (7mm);
\draw[line width=.5mm] (10,2) circle (7mm);
\draw[line width=.25mm, ->] (-0.5,1.8)--(.05,1.8);
\draw[line width=.25mm] (0,1.8)--(1.35,1.8);
\draw[line width=.25mm] (-0.5,2.2)--(0,2.2);
\draw[line width=.25mm,<-] (-0.05,2.2)--(1.35,2.2);
\draw[line width=.25mm, ->] (2.66,2.2)--(4.05,2.2);
\draw[line width=.25mm] (4,2.2)--(5.35,2.2);
\draw[line width=.25mm] (2.66,1.8)--(4,1.8);
\draw[line width=.25mm,<-] (3.95,1.8)--(5.35,1.8);
\draw[line width=.25mm, ->] (6.66,1.8)--(8.05,1.8);
\draw[line width=.25mm] (8,1.8)--(9.35,1.8);
\draw[line width=.25mm] (6.66,2.2)--(8,2.2);
\draw[line width=.25mm,<-] (7.95,2.2)--(9.35,2.2);
\draw[line width=.25mm, ->] (10.66,2.2)--(12.05,2.2);
\draw[line width=.25mm] (12,2.2)--(12.5,2.2);
\draw[line width=.25mm] (10.66,1.8)--(12,1.8);
\draw[line width=.25mm,<-] (11.95,1.8)--(12.5,1.8);
\draw (2,2) node  []  {$A_{I-1}$};
\draw (6,2) node  []  {$A_{I}$};
\draw (10,2) node  []  {$A_{I+1}$};
\draw (2,1) node  []  {$k$};
\draw (6,1) node  []  {$-k$};
\draw (10,1) node  []  {$k$};
\draw (0,2.6) node  []  {$\bar{q}_{I-2,1}$};
\draw (0,1.4) node  []  {$q^2_{I-2}$};
\draw (4.05,2.6) node  []  {$\tilde{q}_{I-1,\dot{1}}$};
\draw (4.05,1.4) node  []  {$\bar{\tilde{q}}^{\dot{2}}_{I-1}$};
\draw (8,2.6) node  []  {$\bar{q}_{I,1}$};
\draw (8,1.4) node  []  {$q^2_{I}$};
\draw (12,2.6) node  []  {$\tilde{q}_{I+1,\dot{1}}$};
\draw (12,1.4) node  []  {$\bar{\tilde{q}}^{\dot{2}}_{I+1}$};
\end{tikzpicture}
\caption{The decomposition of the $\cN=4$ matter multiplets into pairs of chiral multiplets. Only the chiral scalar of each multiplet is indicated explicitly.}
\label{fig:N=4/2quiver}
\end{figure}
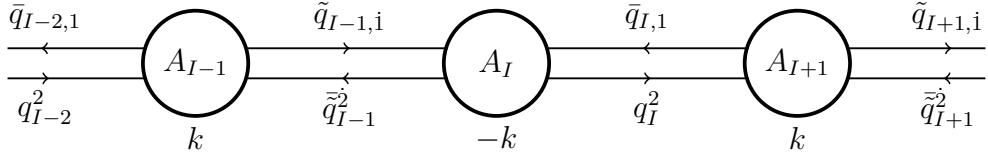

The (twisted) hyper multiplets can be decomposed into pairs of chiral multiplets. 
Figure~\ref{fig:N=4/2quiver} shows the chiral scalar in this decomposition explicitly. Here and throughout chiral fields 
are denoted as solid arrows and when needed, the anti-chiral fields are represented by dashed arrows. 
The orientation of the arrows stands for the field's representation. For example, the fields $q_I^2$ is in the 
$(\Box,\bar\Box)$ of $U(N_I)\times U(N_{I+1})$ and $\bar{q}_{I\,1}$ is in the conjugate representation.

The $SU(2)_L\times SU(2)_R$ R-symmetry indices $a,b=1,2$ and $\dot a,\dot b=\dot 1,\dot 2$ are raised and lowered using the appropriate epsilon symbols: $v^a=\epsilon^{ab}v_ b$ and $v_a=\epsilon_{ab}v^b$ with $\epsilon^{12}=\epsilon_{21}=1$, and similarly for the dotted indices.

We define moment maps and currents, which are bilinears of the hypermultiplets and twisted hypermultiplets in the adjoint representation of $U(N_I)$, as follows
\bal
\mu_{I}{}^a_{\ b}&=q_I^a\bar q_{I\,b}-\frac{1}{2}\delta^a_bq_I^c\bar q_{I\,c}\,,
\qquad&
j_I^{a\dot b}&=q_I^a\bar\psi_I^{\dot b}-\epsilon^{ac}\epsilon^{\dot b\dot c}\psi_{I\,\dot c}\bar q_{I\,c}\,,
\\
\tilde\mu_{I}{}^{\dot a}_{\ \dot b}&=\bar{\tilde q}_{I-1}^{\,\dot a}\tilde q_{I-1\,\dot b}
-\frac{1}{2}\delta^{\dot a}_{\dot b}\bar{\tilde q}_{I-1}^{\,\dot c}\tilde q_{I-1\,\dot c}\,,
\qquad&
\tilde\jmath_I^{\,\dot ba}&=\bar{\tilde q}_{I-1}^{\,\dot b}\tilde \psi_{I-1}^{a}
-\epsilon^{\dot b\dot c}\epsilon^{ac}\bar{\tilde\psi}_{I-1\,c}\tilde q_{I-1\,\dot c}\,,
\\
\nu_{I}&=q_I^a\bar q_{I\,a}\,,
\qquad&
\tilde\nu_{I}&=\bar{\tilde q}_{I-1}^{\,\dot a}\tilde q_{I-1\,\dot a}\,.
\eal
Similar bilinears (with the appropriate replacement of hypermultiplets and twisted hypermultiplets) exist also for the other nodes. For example, for the $I+1$ node one can define $\nu_{I+1}=\bar q_{Ia}q_I^a$. Note that in this notation the index of the moment maps represents the node under which they are charged, rather than the fields they are made of, as is the case in the notation of \cite{Imamura:2008dt,Cooke:2015ila}. In particular, $\mu_{I+1}{}_a^{\ b}$ is made of the same fields $\bar q_{I\,a}$ and $q_I^b$ as $\mu_{I}{}^b_{\ a}$, but it is charged under a different group. 
The moment maps are triplets of the respective $SU(2)$ R-symmetry group and 
are used below to construct the basic bosonic Wilson loops. The moment maps can be thought of as 
the generalization to ${\cal N}=4$ of the ${\cal N}=2$ scalar $\sigma$, though the 
latter is an auxiliary field in an off-shell formulation while we work in an on-shell formulation. 
We provide some details on this correspondence in Appendix~\ref{app:susytransformations}.

As stated in the Introduction, we define the theory on $S^3$ and the Wilson loops we construct are supported along the equator of this sphere. The corresponding ${\cal N}=4$ supersymmetry transformations are given in the appendix in (\ref{SUSY2}), along with details about how they relate to the ones for the ${\cal N}=2$ theory \cite{Hama:2010av,asano2012large}.


\section{Single-node Wilson  loops}
\label{sec:bos}

We start by constructing Wilson loops coupling to a single gauge field and hence suppress the $I$ index on the fields. The most symmetric such loop is a circle coupling to both the untwisted and twisted moment maps through the connection 
\beq
\label{1/4}
\cA=A_\varphi-\frac{i}{k}(\mu^1_{\ 1}-\mu^2_{\ 2}+\tilde\mu^{\dot1}_{\ \dot1}-\tilde\mu^{\dot2}_{\ \dot2}).
\eeq
This choice of scalar coupling can be motivated by considering what is the natural generalization  to the ${\cal N}=4$ case of the scalar $\sigma$ appearing in the Wilson loop in ${\cal N}=2$ Chern-Simons-matter theory \cite{Gaiotto:2007qi}. Since the moment maps are triplets of the R-symmetry group for ${\cal N}=4$, the coupling above corresponds to picking the moment maps along the `third' direction of the R-symmetry triplet, so can be denoted alternatively as $\mu^3$ and $\tilde\mu^3$.

From (\ref{SUSY2}) one finds the supersymmetry transformations of the moment maps
\beq
\label{deltamu}
\delta \mu_I{}^a_{\ b}=\xi_{b\dot c}j_I^{a\dot c}-\frac{1}{2}\delta^a_b\xi_{c\dot c}j_I^{c\dot c}\,,\qquad
\delta\tilde\mu_I{}^{\dot a}_{\ \dot b}=-\xi_{c\dot b}\tilde\jmath_{I}^{\,\dot ac}
+\frac{1}{2}\delta^{\dot a}_{\dot b}\xi_{c\dot c}\tilde\jmath_{I}^{\,\dot c c}\,,
\eeq
so the variation of the connection in (\ref{1/4}) is
\beq
\label{var1/4}
\delta\cA=\frac{i}{k}\xi_{a\dot b}\gamma_\varphi(j^{a\dot b}-\tilde\jmath^{\,\dot ba})
-\frac{i}{k}(\xi_{1\dot b}j^{1\dot b}-\xi_{2\dot b}j^{2\dot b}
-\xi_{a\dot1}\tilde\jmath^{\dot1 a}+\xi_{a\dot2}\tilde\jmath^{\dot2 a})\,,
\eeq
with $\gamma_\varphi$ along the equatorial circle on which the loop is supported. There are no solutions to $\delta\cA=0$ for $\xi_{1\dot2}$ and $\xi_{2\dot1}$, while for the other components one finds the conditions
\beq
\label{killingspinors1/4}
\xi_{1\dot1}(1-\gamma_\varphi)=\xi_{2\dot2}(1+\gamma_\varphi)=0\,.
\eeq
The resulting Wilson loop is hence $1/4$ BPS, and it is in fact the same as the $1/2$ BPS loop in $\cN=2$ theories \cite{Gaiotto:2007qi} or the $1/6$ BPS loop in ABJ(M) theory, as in \cite{Drukker:2008zx, Chen:2008bp,Rey:2008bh}. 

To be more explicit, each supersymmetry parameter $\xi_{a\dot b}$ is a linear combination of four 
Killing-spinors on $S^3$. We label them as $\xi^l$, $\xi^{\bar l}$, $\xi^r$, $\xi^{\bar r}$, according to their chiralities.
They obey
\beq
\label{xi}
\nabla_\mu \xi^{l,\bar l} = \frac{i}{2}\gamma_\mu \xi^{l, \bar l}\,, \qquad 
\nabla_\mu \xi^{r, \bar r} = - \frac{i}{2}\gamma_\mu \xi^{r, \bar r}\,.
\eeq
Along the circle we take
$\gamma_\varphi=\sigma_3$ and these reduce to \cite{Assel:2015oxa}
\beq
\label{killingspinors}
\xi^l_\alpha=\begin{pmatrix}1\\0\end{pmatrix},
\qquad
\xi^{\bar l}_\alpha=\begin{pmatrix}0\\1\end{pmatrix},
\qquad
\xi^r_\alpha=\begin{pmatrix}e^{-i\varphi}\\0\end{pmatrix},
\qquad
\xi^{\bar r}_\alpha=\begin{pmatrix}0\\e^{i\varphi}\end{pmatrix}.
\eeq

Equation \eqref{killingspinors1/4} restricts $\xi_{1\dot 1}$ 
to the two chiralities $\xi^{\bar l}$ and $\xi^{\bar r}$, while $\xi_{2\dot2}$ is a linear combination of $\xi^l$ and $\xi^r$. 
We can write the corresponding four supersymmetries as
\beq
\label{4supercharges}
Q^{\dot11}_{\bar l}\,,\quad
Q^{\dot11}_{\bar r}\,,\quad
Q^{\dot22}_{l}\,,\quad
Q^{\dot22}_{r}\,.
\eeq

In the subsequent sections we construct families of Wilson loop operators that preserve all
or particular linear combinations of those supercharges. These loops couple to two or more nodes of the quiver and involve combining several gauge connections and couplings to more bifundamental fields. However, there is still a family of Wilson loops involving just a single 
node that preserve only two supercharges, being therefore $1/8$ BPS. Following the logic of 
the $1/4$ BPS Wilson loops in $\cN=4$ SYM in 4d \cite{Drukker:2006ga}, the connection (\ref{1/4}) 
is naturally generalized introducing a ``latitude''%
\footnote{As already mentioned in the Introduction, note that this is only a latitude in the internal space of scalar couplings, while the loop is still an equator of the $S^3$.}
 angle $\theta$ and an azimuthal angle $\varphi_0$, as follows
\beq
\label{1/8}
\cA^{\theta}=A_\varphi
-\frac{i}{k}\left(\mu^1_{\ 1}-\mu^2_{\ 2}
+\cos\theta\,(\tilde\mu^{\dot1}_{\ \dot1}-\tilde\mu^{\dot2}_{\ \dot2})
+\sin\theta (e^{-i{(\varphi-\varphi_0)}}\tilde\mu^{\dot1}_{\ \dot2}
+e^{i(\varphi-\varphi_0)}\tilde\mu^{\dot2}_{\ \dot1})\right)
\,.
\eeq
Notice that only the couplings to the twisted hypermultiplets are modified, via their moment maps. 
This operator is now coupled to three different moment maps, which can alternatively be written as $\tilde\mu^3$ and $\tilde\mu^\pm$. This deformation is not possible in theories with only $\cN=2$ symmetry, with only one $\sigma$ field in the vector multiplet.
An analogous construction where 
the untwisted moment maps are modified and the twisted ones remain as in \eqref{1/4} also works, 
but deforming both at the same time does not give BPS operators. 
This deformation involves two parameters, $\theta$ and $\varphi_0$, which define an $S^2$. For simplicity we set $\varphi_0=0$ in the following.

The supersymmetry variation of \eqref{1/8} gives
\bal
\delta\cA^{\theta}&=\frac{i}{k}\left(\xi_{a\dot b}\gamma_\varphi(j^{a\dot b}-\tilde\jmath^{\,\dot ba})
-(\xi_{1\dot b}j^{1\dot b}-\xi_{2\dot b}j^{2\dot b})
\right.\\&\quad{}\left. \hskip 1cm
+\cos\theta\,(\xi_{a\dot1}\tilde\jmath^{\,\dot1a}-\xi_{a\dot2}\tilde\jmath^{\,\dot2a})+\sin\theta (e^{-i\varphi}\xi_{a\dot 2}\tilde\jmath^{\,\dot 1a}
+e^{i\varphi}\xi_{a\dot 1}\tilde\jmath^{\,\dot 2a})\right)\,.
\eal
Requiring this to vanish and collecting terms according to the components of the currents, one finds conditions on the supersymmetry parameters which are all solved imposing
\bal
\label{susy-cond}
&\xi_{1\dot a}(1-\gamma_\varphi)
=\xi_{2\dot a}(1+\gamma_\varphi)=0\,,
\cr
&\xi_{a\dot1}(\gamma_\varphi -\cos\theta)=\sin\theta\,e^{-i\varphi}\xi_{a\dot2}\,.
\eal

The first line of \eqref{susy-cond} sets 
$\xi_{1\dot a}^{l}=\xi_{1\dot a}^{r}=\xi_{2\dot a}^{\bar l}=\xi_{2\dot a}^{\bar r}=0$. The second line also eliminates $\xi_{a\dot 2}^{r}=\xi_{a\dot 2}^{\bar l}=\xi_{a\dot 1}^{l}=\xi_{a\dot 1}^{\bar r}=0$. The remaining  $\xi_{1\dot 1}^{\bar l}$, $\xi_{1\dot 2}^{\bar r}$, $\xi_{2\dot 1}^r$, $\xi_{2\dot 2}^l$ are related by
\beq
\label{killing1}
\xi_{1\dot1}^{\bar l}=\frac{\sin\theta}{1-\cos\theta}\,\xi_{1\dot 2}^{\bar r}=\cot\frac{\theta}{2}\,\xi_{1\dot 2}^{\bar r}\,,
\qquad
\xi_{2\dot1}^{r}=-\frac{\sin\theta}{1+\cos\theta}\,\xi_{2\dot 2}^l=-\tan\frac{\theta}{2}\,\xi_{2\dot 2}^{l}\,.
\eeq
We find finally the two independent supercharges preserved by the loop
\beq
\label{supercharges}
\cQ_1=\cos\frac{\theta}{2}\,Q^{\dot 11}_{\bar l}+\sin\frac{\theta}{2}\,Q^{\dot21}_{\bar r}\,,
\qquad
\cQ_2=\cos\frac{\theta}{2}\,Q^{\dot 22}_{l}-\sin\frac{\theta}{2}\,Q^{\dot12}_{r}\,,
\eeq
which is then $1/8$ BPS, as advertised. Notice that the supercharges in \eqref{supercharges} are not a subset of those in \eqref{4supercharges}. The reason for this is that we kept the circle on the equator of $S^3$. Were we to follow the logic of \cite{Cardinali:2012ru} and place the loop at a latitude angle of $\pi/2-\theta$, which can be done via a conformal transformation on the $S^3$, the resulting loops would preserve $Q^{\dot 11}_{\bar l}$ and $Q^{\dot 22}_{l}$, which are indeed a subset of \eqref{4supercharges}.


\section{Hyperloops at $\theta=0$}
\label{sec:theta=0}

In this and the next section we construct what we dub {\it hyperloops}: BPS Wilson loops involving multiple gauge fields and couplings to the hypermultiplets beyond their bilinears.
Viewed in $\cN=2$ language, these loops were already constructed in \cite{drukker2020bps} (large subclasses of them were previously found in \cite{Ouyang:2015iza}), so this section is mostly a review of that paper, specializing to $\cN=4$ theories. In addition to explaining 
the structure of the loops, we also review the moduli space of BPS loops and its relation to 
quiver varieties.

The analysis of \cite{drukker2020bps} gives families of Wilson loops preserving either all the supercharges 
in \eqref{4supercharges} or one of the linear combinations
\beq
\cQ^L_+=Q^{\dot11}_{\bar l}+Q^{\dot22}_{l}\,,
\qquad
\cQ^L_-=Q^{\dot11}_{\bar l}-Q^{\dot22}_{l}\,.
\eeq

The possible hyperloops can be visualized by quiver diagrams, which may include some or all of the nodes 
and edges of the original quiver defining the gauge theory. Let us review the connection between quiver diagrams and 
hyperloops. First, we include a node for each vector multiplet the loop couples to, solid arrows for the 
chiral fields and dashed arrows for the anti-chirals. Some of the nodes are denoted by squiggly 
circles and some by unsquiggly ones. This represents that the connection of the 
gauge field in the squiggly nodes has an extra shift by%
\footnote{Here and throughout we set the radius of the sphere to $R=1$. 
For a generic radius the shifts would be $\pm \frac{1}{2R}$, and similarly for the shifts in the next section.}  
$1/2$,
and supersymmetry requires alternating squiggly and unsquiggly nodes. 
Each node is decorated by integers $p_I$, indicating the multiplicity of the gauge field in the 
hyperloop. We indicate the chiral fields coupling to the hyperloop 
(according to the decomposition in Figure~\ref{fig:N=4/2quiver}) by solid arrows and 
the anti-chirals by dashed arrows. Arrows between the $I$ and $I+1$ nodes are decorated 
by $p_I\times p_{I+1}$ complex parameters $\alpha_I$ and $\bar\alpha_I$ for anti-chirals, 
though they are not complex conjugates.

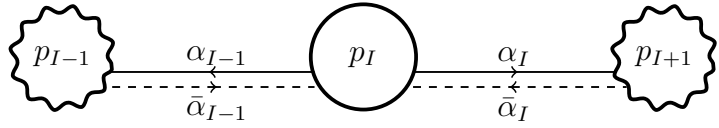
\begin{figure}[H]
\centering
\begin{tikzpicture}
\draw[decoration={snake,amplitude = .5mm,segment length=3.46mm},decorate, line width=.5mm] (2,1.9) circle (6.5mm);
\draw[line width=.5mm] (6,2) circle (7mm);
\draw[decoration={snake,amplitude = .5mm,segment length=3.46mm},decorate, line width=.5mm] (10,1.9) circle (6.5mm);
\draw[line width=.25mm, dashed, ->] (2.66,1.6)--(4.05,1.6);
\draw[line width=.25mm,dashed] (4,1.6)--(5.35,1.6);
\draw[line width=.25mm] (2.6,1.8)--(4,1.8);
\draw[line width=.25mm,<-] (3.95,1.8)--(5.35,1.8);
\draw[line width=.25mm, ->] (6.66,1.8)--(8.05,1.8);
\draw[line width=.25mm] (8,1.8)--(9.35,1.8);
\draw[line width=.25mm,dashed, ] (6.66,1.6)--(8,1.6);
\draw[line width=.25mm,dashed, <-] (7.95,1.6)--(9.5,1.6);
\draw (2,2) node  []  {$p_{I-1}$};
\draw (6,2) node  []  {$p_{I}$};
\draw (10,2) node  []  {$p_{I+1}$};
\draw (4.05,2) node  []  {$\alpha_{I-1}$};
\draw (4.05,1.3) node  []  {$\bar{\alpha}_{I-1}$};
\draw (8,2) node  []  {${\alpha}_{I}$};
\draw (8,1.3) node  []  {$\bar\alpha_{I}$};
\end{tikzpicture}
\caption{A quiver diagram for a $1/4$ BPS Wilson loop.}
\label{fig:1/4-quiver}
\end{figure}

These diagrams represent $1/4$ BPS hyperloops, preserving all four supercharges in 
\eqref{4supercharges} if all solid arrows point into squiggly nodes.%
\footnote{In a different gauge, they all point out of the squiggly nodes.} 
If the decorated quiver contains solid arrows pointing both in and out of the 
squiggly circles, the loop preserves only one of the supercharges in 
\eqref{4supercharges}, either $\cQ_+^{L}$ or $\cQ_-^{L}$, and is $1/16$ BPS.

We begin by describing the $1/4$ BPS loops. 
Given a decorated quiver diagram, the hyperloop is constructed as follows. One first 
assembles a diagonal matrix $\cL_0$, with $p_I$ copies of the connection $\cA_I$ \eqref{1/4} 
for every unsquiggly node. For squiggly nodes the connection is augmented to $\cA_I+1/2$.
For example, let us consider the quiver in Figure~\ref{fig:1/4-quiver} with $p_{I-1}=p_{I+1}=1$ and $p_I=2$.
The corresponding $\cL_0$ matrix is
\beq
\label{cL0-1}
\cL_0=\begin{pmatrix}
\cA_{I-1}+\frac{1}{2}&0&0&0\\
0&\ \cA_{I}\ &0&0\\
0&0&\ \cA_{I}\ &0\\
0&0&0&\cA_{I+1}+\frac{1}{2}
\end{pmatrix}.
\eeq
Next we define the matrix
\beq
\label{G1}
\cG=\begin{pmatrix}
0&
\ \bar\alpha^1_{I-1}\tilde q_{I-1\,\dot2}\ &\ \bar\alpha^2_{I-1}\tilde q_{I-1\,\dot2}\
&0\\
\alpha_{I-1}^1\bar{\tilde q}^{\dot2}_{I-1}&\ 0\ &\ 0\ &\alpha_I^1q_I^2\\
\alpha_{I-1}^2\bar{\tilde q}^{\dot2}_{I-1}&0&0&\alpha_I^2q_I^2\\
0&
\bar\alpha^1_I\bar q_{I\,2}&\bar\alpha^2_I\bar q_{I\,2}
&0
\end{pmatrix},
\eeq
from which we construct the superconnection (see for example \cite{Drukker:2019bev}) 
\beq
\label{superconn}
\cL_{\alpha,\bar\alpha} = \cL_0+i\cQ_+^L\cG+\cG^2\,.
\eeq
The hyperloop is then
\beq
W_{\alpha,\bar\alpha} = \sTr\cP\exp\left[i \oint
\cL_{\alpha,\bar\alpha}\,|\dot x|\,ds\right].
\eeq
The supertrace treats the unsquiggly nodes ($I$) as even and the squiggly ones ($I\pm1$) as odd.

\begin{figure}[H]
\centering
\begin{tikzpicture}
\draw[decoration={snake,amplitude = .5mm,segment length=3.46mm},decorate, line width=.5mm] (6,1.9) circle (6.5mm);
\draw[line width=.5mm] (2,2) circle (7mm);
\draw[line width=.5mm] (10,2) circle (7mm);
\draw[line width=.25mm, ->] (2.65,2.25)--(4.05,2.25);
\draw[line width=.25mm] (4,2.25)--(5.45,2.25);
\draw[line width=.25mm,dashed] (2.55,2.45)--(4,2.45);
\draw[line width=.25mm,dashed,<-] (3.95,2.45)--(5.45,2.45);
\draw[line width=.25mm] (6.6,2.25)--(8,2.25);
\draw[line width=.25mm,<-] (7.95,2.25)--(9.35,2.25);
\draw[line width=.25mm, dashed,->] (6.55,2.45)--(8.05,2.45);
\draw[line width=.25mm,dashed ] (8,2.45)--(9.45,2.45);
\draw (2,2) node  []  {$p_{I-1}$};
\draw (6,2) node  []  {$p_{I}$};
\draw (10,2) node  []  {$p_{I+1}$};
\draw (4.5,2) node  []  {$\beta_{I-1}$};
\draw (4.5,2.7) node  []  {$\bar{\beta}_{I-1}$};
\draw (8.5,2) node  []  {${\beta}_{I}$};
\draw (8.5,2.7) node  []  {$\bar\beta_{I}$};
\end{tikzpicture}
\caption{The other quiver diagram giving another family of $1/4$ BPS hyperloops.}
\label{fig:1/4-quiver2}
\end{figure}
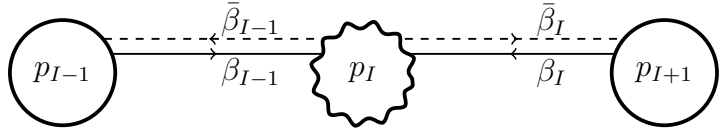

Another branch of 1/4 BPS hyperloops comes from swapping squiggly and unsquiggly nodes 
as in in Figure~\ref{fig:1/4-quiver2}, with the same values of the $p_I$'s as above.
In this case, the bosonic connection is\footnote{To avoid cluttering the notation too much, we denote the bosonic connections and the matrices $\cG$ of all these examples with the same symbols: $\cL_0$ and $\cG$. We always refer to explicit equations, so this should hopefully not lead to confusion. }
\beq
\label{cL0-2}
\cL_0=\begin{pmatrix}
\cA_{I-1}&0&0&0\\
0&\cA_{I}+\frac{1}{2}&0&0\\
0&0&\cA_{I}+\frac{1}{2}&0\\
0&0&0&\cA_{I+1}
\end{pmatrix}
\eeq
and
\beq
\label{G2}
\cG=\begin{pmatrix}
0&
\ \beta^1_{I-1}\tilde q_{I-1\,\dot1}\ &\ \beta^2_{I-1}\tilde q_{I-1\,\dot1}\
&0\\
\bar\beta_{I-1}^1\bar{\tilde q}^{\dot1}_{I-1}&\ 0\ &\ 0\ &\bar\beta_I^1q_I^1\\
\bar\beta_{I-1}^2\bar{\tilde q}^{\dot1}_{I-1}&0&0&\bar\beta_I^2q_I^1\\
0&
\beta^1_I\bar q_{I\,1}&\beta^2_I\bar q_{I\,1}
&0
\end{pmatrix}.
\eeq
The rest of the construction is exactly as above with a superconnection and hyperloop that could be denoted as $\cL_{\beta,\bar\beta}$ and $W_{\beta,\bar\beta}$ (with the parameters $\beta$ and $\bar\beta$ being complex, but, again, not complex conjugates of each other).

\begin{figure}[H]
\centering
\begin{tikzpicture}
\draw[decoration={snake,amplitude = .5mm,segment length=3.46mm},decorate, line width=.5mm] (2,1.9) circle (6.5mm);
\draw[line width=.5mm] (6,2) circle (7mm);
\draw[decoration={snake,amplitude = .5mm,segment length=3.46mm},decorate, line width=.5mm] (10,1.9) circle (6.5mm);
\draw[line width=.25mm, ->] (2.62,2.25)--(4.05,2.25);
\draw[line width=.25mm] (4,2.25)--(5.35,2.25);
\draw[line width=.25mm,dashed] (2.5,2.45)--(4,2.45);
\draw[line width=.25mm,dashed,<-] (3.95,2.45)--(5.45,2.45);
\draw[line width=.25mm] (2.55,1.75)--(4,1.75);
\draw[line width=.25mm,<-] (3.95,1.75)--(5.35,1.75);
\draw[line width=.25mm, dashed, ->] (2.66,1.55)--(4.05,1.55);
\draw[line width=.25mm,dashed] (4,1.55)--(5.35,1.55);
\draw[line width=.25mm, ->] (6.66,1.75)--(8.05,1.75);
\draw[line width=.25mm] (8,1.75)--(9.35,1.75);
\draw[line width=.25mm,dashed, ] (6.66,1.55)--(8,1.55);
\draw[line width=.25mm,dashed, <-] (7.95,1.55)--(9.5,1.55);
\draw[line width=.25mm] (6.66,2.25)--(8,2.25);
\draw[line width=.25mm,<-] (7.95,2.25)--(9.45,2.25);
\draw[line width=.25mm, dashed,->] (6.6,2.45)--(8.05,2.45);
\draw[line width=.25mm,dashed ] (8,2.45)--(9.45,2.45);
\draw (2,2) node  []  {$p_{I-1}$};
\draw (6,2) node  []  {$p_{I}$};
\draw (10,2) node  []  {$p_{I+1}$};
\draw (3.5,2) node  []  {$\alpha_{I-1}$};
\draw (3.5,1.3) node  []  {$\bar{\alpha}_{I-1}$};
\draw (4.5,2) node  []  {$\beta_{I-1}$};
\draw (4.5,2.7) node  []  {$\bar{\beta}_{I-1}$};
\draw (7.5,2) node  []  {${\alpha}_{I}$};
\draw (7.5,1.3) node  []  {$\bar\alpha_{I}$};
\draw (8.5,2) node  []  {${\beta}_{I}$};
\draw (8.5,2.7) node  []  {$\bar\beta_{I}$};
\end{tikzpicture}
\caption{A quiver diagram for a $1/16$ BPS} hyperloop.
\label{fig:1/16-quiver}
\end{figure}
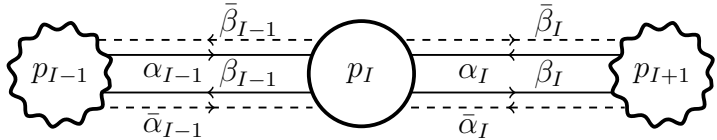

The $1/16$ BPS loops are obtained by couplings to all the chiral and anti-chiral fields with parameters 
$\alpha_I,\bar{\alpha}_I$ and $\beta_I,\bar{\beta}_I$. This can be achieved by a gauge transformation of 
$\cL_0$ in \eqref{cL0-2} to the form in \eqref{cL0-1}, which corresponds to exchanging squiggly and 
unsquiggly nodes in Figure \ref{fig:1/4-quiver2}. This results in the extra phases
\beq
\label{gwithphases}
\cG\to\begin{pmatrix}
0&
\ e^{i\varphi}\beta^1_{I-1}\tilde q_{I-1\,\dot1}\ &\ e^{i\varphi}\beta^2_{I-1}\tilde q_{I-1\,\dot1}\
&0\\
e^{-i\varphi}\bar\beta_{I-1}^1\bar{\tilde q}^{\dot1}_{I-1}&\ 0\ &\ 0\ &e^{-i\varphi}\bar\beta_I^1q_I^1\\
e^{-i\varphi}\bar\beta_{I-1}^2\bar{\tilde q}^{\dot1}_{I-1}&0&0&e^{-i\varphi}\bar\beta_I^2q_I^1\\
0&
e^{i\varphi}\beta^1_I\bar q_{I\,1}&e^{i\varphi}\beta^2_I\bar q_{I\,1}
&0
\end{pmatrix}.
\eeq
We add this to the expression in \eqref{G1} to find a new $\cG$ given by
\bal
\label{16GbarG}
\begin{pmatrix}
0& \hskip-1cm \bar\alpha^1_{I-1}\tilde q_{I-1\,\dot2}+e^{i\varphi}\beta^1_{I-1}\tilde q_{I-1\,\dot1} \ &\  \bar\alpha^2_{I-1}\tilde q_{I-1\,\dot2}+e^{i\varphi}\beta^2_{I-1}\tilde q_{I-1\,\dot1} &\hskip -1cm 0
\\
\alpha_{I-1}^1\bar{\tilde q}^{\dot2}_{I-1}+e^{-i\varphi}\bar\beta_{I-1}^1\bar{\tilde q}^{\dot1}_{I-1} &\hskip -1cm  0 & 0 &\hskip -1cm \alpha_I^1q_I^2++e^{-i\varphi}\bar\beta_I^1q_I^1
\\
\alpha_{I-1}^2\bar{\tilde q}^{\dot2}_{I-1}+e^{-i\varphi}\bar\beta_{I-1}^2\bar{\tilde q}^{\dot1}_{I-1}&\hskip -1cm 0&0&\hskip -1cm \alpha_I^2q_I^2+e^{-i\varphi}\bar\beta_I^2q_I^1
\\
0&\hskip -1cm \bar\alpha^1_I\bar q_{I\,2}+e^{i\varphi}\beta^1_I\bar q_{I\,1}&\bar\alpha^2_I\bar q_{I\,2}+e^{i\varphi}\beta^2_I\bar q_{I\,1}&\hskip -1cm 0
\end{pmatrix}.
\eal
The rest of the construction follows as before. The superconnection can be constructed with $\cQ_+^{L}$ as in (\ref{superconn}) or, alternatively, with $\cQ_-^{L}$, leading to two different classes of $1/16$ BPS hyperloops, graphically represented as in Figure~\ref{fig:1/16-quiver}.

For any quiver diagram we can always exchange squiggly and unsquiggly nodes ({\it i.e.} grading) at the price of adding phases to the matter fields, as in \eqref{gwithphases}. This leads to the same loop operator, but in a different gauge. Note that in the case of $1/16$ BPS loops shown in Figure \ref{fig:1/16-quiver}, the two gradings of the quiver are gauge equivalent. This is not the case for families of $1/4$ BPS loops, where the two gradings also represent which of the chiral fields are included, as illustrated in Figures \ref{fig:1/4-quiver} and \ref{fig:1/4-quiver2}.


\subsection{Moduli spaces}
\label{subsec:modulispaces}

The hyperloops we construct are built upon a connection $\cL_0$ to which we add bifundamental 
couplings encoded in the matrix $\cG$. The deformation is thus described by the parameters in 
this matrix, which are a set of complex numbers $\alpha$, $\bar\alpha$, etc. This description is 
however redundant, so the moduli space of hyperloops is a quotient of the space of these 
parameters, as described below.

Wilson loops are intimately related to gauge invariance, and in particular are gauge invariant observables. 
The invariance under $\prod_I U(N_I)$ gauge transformations are built into the definition of the 
hyperloops, but they posses a larger (global) gauge symmetry. $\cL$ can be thought 
of as a superconnection, which in the example above is in the superalgebra 
$GL(N_{I-1}+N_{I+1}|2N_{I})$. 
We do not expect local-gauge symmetry under this 
group, as it is not a symmetry of the theory, but constant gauge transformations 
do not require extra gauge fields. Of those, the transformations that preserve our formulation and 
are not the gauge symmetries of the quiver theory are the centralizer of $\cL_0$, which in our examples 
is
\beq
(\bC^*)^2\times GL(2,\bC)\,.
\eeq
Concretely this symmetry acts by conjugation
\beq
\cL_0 \rightarrow
\begin{pmatrix}
x^{-1} & \begin{matrix}
0 & 0
\end{matrix} & 0\\
\begin{matrix}
0 \\ 0
\end{matrix} & S^{-1} & \begin{matrix}
0 \\ 0
\end{matrix}\\
0 & \begin{matrix}
0 & 0
\end{matrix} & y^{-1}
\end{pmatrix}
\cL_0
\begin{pmatrix}
x & \begin{matrix}
0 & 0
\end{matrix} & 0\\
\begin{matrix}
0 \\ 0
\end{matrix} & S & \begin{matrix}
0 \\ 0
\end{matrix}\\
0 & \begin{matrix}
0 & 0
\end{matrix} & y
\end{pmatrix}
=\cL_0\,,
\qquad S\in GL(2,\bC)\,.
\eeq
The action on $\cG$ and consequently on $\cL_{\alpha,\bar\alpha}$ is generally non-trivial, 
mapping superconnections with different parameters to each-other. Note though that 
matrices in the center of $GL(4,\bC)$, {\it i.e.} where $y=x$ and $S=\diag(x,x)$ do commute with 
all matrices, so should be excluded, thus the gauge symmetry is really
\beq
\label{symmofL}
S((\bC^*)^2\times GL(2,\bC))\,.
\eeq

Remaining with our examples above, we describe the $1/4$ BPS loops with eight complex parameters, 
see \eqref{G1} and \eqref{G2}, and the $1/16$ BPS loops with sixteen, see \eqref{16GbarG}. 
As argued, this means that the moduli space of $1/4$ BPS loops correspond to two copies of
\beq
\bC^8/\!\!/S((\bC^*)^2\times GL(2,\bC))\,,
\eeq
while for $1/16$ BPS operators we have for either $\cQ^L_+$ or $\cQ^L_-$ a single copy of
\beq
\bC^{16}/\!\!/S((\bC^*)^2\times GL(2,\bC))\,.
\eeq
These are 3- and 11-(complex) dimensional conical spaces, respectively.

These spaces are the usual quiver varieties associated to the quiver representations in 
Figures~\ref{fig:1/4-quiver}, \ref{fig:1/4-quiver2} and~\ref{fig:1/16-quiver}, 
see \cite{Nakajima1999lectures, Hanany:1996ie, Crawley-Boevey, ginzburg2009lectures, kirillov2016quiver}.
We use the double slash notation mirroring the concept of geometric invariant 
theory, in order to point out that we need to 
be careful when identifying the singular orbits of the resulting manifolds. 
In particular, hyperloops with off-diagonal components that are exclusively upper or 
lower triangular are identical as quantum operators for all values of $\alpha$ 
and therefore should be identified.

The analysis can be carried out for more general quivers as follows. In the case of $1/4$ BPS loops, 
each edge between nodes of multiplicities $p_I$ and $p_{I+1}$ has $2p_I p_{I+1}$ complex parameters.
For $1/16$ BPS operators, this becomes $4p_I p_{I+1}$. From these we need to remove the symmetries, 
which amount to a factor of $GL(p_I,\bC)$ for each node with multiplicity $p_I$, apart for the trivial 
action of the center, as explained above \eqref{symmofL}.

Therefore, for a linear quiver of length $L$ we find the moduli space of $1/4$ BPS loops to be two copies of
\beq
\label{open-quiver-moduli}
\bC^{2p_1p_2}\times
\bC^{2p_2p_3}\times\cdots\times\bC^{2p_{L-1}p_L}/\!\!/
S(GL(p_1)\times GL(p_2)\times\cdots\times GL(p_L))\,.
\eeq
For the $1/16$ BPS loops, we loose the second copy and we have the same as above where the 2s in the exponents become 4s.

For a circular quiver, the moduli space of $1/4$ BPS loops is two copies of
\beq
\label{ABJ(M)-moduli}
\bC^{2p_1p_2}\times
\bC^{2p_2p_3}\times\cdots\times\bC^{2p_{L-1}p_L}\times\bC^{2p_{L}p_1}/\!\!/
S(GL(p_1)\times GL(p_2)\times\cdots\times GL(p_L))\,.
\eeq
For $1/16$ BPS loops, we again loose the second copy and the manifold is the one above with 2s replaced by 4s.

All these moduli spaces are the quiver varieties associated to the quiver representations that the 
hyperloops furnish. Before moving on to the next section, where we study the $\theta\neq0$ 
versions of these hyperloops, let us specialize to  
the moduli space of 1/6 BPS loops in ABJ(M) theory with $p_1=p_2=1$. As it is a circular two-node 
quiver, we set $L=2$ in \eqref{ABJ(M)-moduli}, giving two copies
\beq
\bC^{2}\times
\bC^{2}/\!\!/
S((\bC^*)\times(\bC^*))=\bC^4 /\!\!/ \bC^*\,.
\eeq
This quotient is the conifold, as already found in \cite{Drukker:2019bev}.


\section{Hyperloops at $\theta\neq0$}
\label{sec:thetanot0}

Having reviewed the formulation of the loops in terms of quiver representations and implemented it 
for theories with $\cN=4$ supersymmetry, we generalize it to loops based on the bosonic connections deformed as in \eqref{1/8}. We provide here much more detail, as most of these loops are novel.

\subsection{Construction}
\label{subsec:construction}

To construct hyperloops with $\theta\neq 0$ we apply the same logic of  
\cite{Drukker:2019bev}, involving a deformation built from the supercharges in 
\eqref{supercharges}. More specifically, we consider the combinations $\cQ_\pm=\cQ_1\pm\cQ_2$. Note that when $\theta=0$ it follows that $\cQ_\pm=\cQ_\pm^L$. The whole procedure relies on being able to write $(\cQ_\pm)^2\cG={\cD}_\varphi\cG$, for some 
matrix of matter fields $\cG$ and an appropriate covariant derivative $\cD_\varphi$. Since
$(\cQ_1)^2=(\cQ_2)^2=0$, this is the same as acting with $\{\cQ_1,\cQ_2\}$ on $\cG$. In Appendix~\ref{app:conventionsanddoubletransformations} we compute the explicit form of this double transformation acting on the untilded scalar fields $q_I^a$ and on a specific spatially-dependent rotation of the tilded ones, defined by
\bea
\label{rotatedscalarfields}
\tilde{r}_{I\pm 1\,\dot 1}\equiv \cos\frac{\theta}{2}\,\tilde{q}_{I\pm 1\,\dot 1}+e^{-i\varphi}\sin\frac{\theta}{2}\,\tilde{q}_{I\pm 1\,\dot 2} \,,\qquad
\tilde{r}_{I\pm 1\,\dot 2}\equiv \cos\frac{\theta}{2}\,\tilde{q}_{I\pm 1\,\dot 2}-e^{i\varphi}\sin\frac{\theta}{2}\,\tilde{q}_{I\pm 1\,\dot 1} \,.
\eea
These combinations are nice because the connection in (\ref{1/8}) can be written compactly in their terms as
\beq
\label{rotated-cA}
\cA^{\theta}_I=A_{\varphi,I}
-\frac{i}{k}({{\mu_I}^1}_{1}-{{\mu_I}^2}_{2}
+\bar{\tilde r}_{I-1}^{\,\dot1}\tilde r_{I-1\,\dot1}-\bar{\tilde r}_{I-1}^{\,\dot2}\tilde r_{I-1\,\dot2})\,.
\eeq
Moreover, one can show that the double transformation acting on $q_I^a$ and $\tilde{r}_{I\pm 1\, \dot a}$ can be recast as a covariant derivative, see (\ref{Q1Q2q}) and (\ref{Q1Q2tildet}). These fields are then the natural ingredients to write down $\cG$.

The connections appearing in the double deformations are as in \eqref{rotated-cA}, with extra 
shifts of $\pm\frac12$ and $\pm\frac12\cos\theta$ that can be viewed as the coupling to a background 
field on the sphere. This is implemented by shifting the original connections, as in the previous section. 
Let us note that the effect of these shifts is to introduce phases like $e^{\pm i\pi\cos\theta}$ 
in the definition of the Wilson loop, 
which are compensated for in the definition of the trace. Recall that in the original formulation of the 
1/2 BPS Wilson loop of \cite{Drukker:2009hy} a trace was required to make them gauge invariant, 
but in the gauge introduced in \cite{Drukker:2019bev} and in Section~\ref{sec:theta=0}, this is 
replaced with a more natural supertrace. For gauge invariance of hyperloops at $\theta\neq0$, the 
definition of supertrace should include  $\pm1$ and $e^{\pm i\pi\cos\theta}$ gradings.
The generic Wilson loop with both $\alpha$ and $\beta$ couplings is then defined as
\beq
W^\theta_{\alpha,\bar\alpha,\beta,\bar\beta} = \sTr_\theta\cP\exp\left[i \oint
\cL^\theta_{\alpha,\bar\alpha,\beta,\bar\beta}\,|\dot x|\,ds\right],
\eeq 
where $\sTr_\theta$ includes the generalized gradings mentioned above to compensate for the 
effect of the shifts.

To be concrete, we illustrate this explicitly in an example of a hyperloop involving 
the three nodes $I-1$, $I$ and $I+1$, as before. For $p_{I-1}=p_I=p_{I+1}=1$, we now have
\beq
\label{Lbos3node}
\cL^\theta_0=\begin{pmatrix}
\cA_{I-1}^\theta + \frac{1}{2}& 0 & 0 \\
0 & \cA_I^\theta & 0 \\
0& 0& \cA_{I+1}^\theta +\frac{1}{2}\cos\theta
\end{pmatrix}\,.
\eeq
The explicit expression for $\cA_{I\pm1}^\theta$ is written in (\ref{AIpm1}). We associate to this choice of connection the quiver diagram in Figure~\ref{fig:1/4-quiver}, though there are now some differences: the loop preserves only two supercharges, so it is $1/8$ BPS; 
the right node has a shift of $\frac12\cos\theta$ instead of $\frac12$ and 
the left arrows are couplings to the rotated fields $\tilde r_{I-1\,\dot 2}$ and their conjugates, rather than to the $\tilde q$'s.
The corresponding hyperloop is
\bal
W^\theta_0=\sTr_\theta \cP \exp \left[i \oint \cL^\theta_0 |\dot x| ds\right]
 &\equiv 
-\Tr \cP \exp \left[i \oint \left(\cA^\theta_{I-1}+\frac{1}{2}\right) |\dot x| ds\right] \\
&\quad{}+ \Tr \cP \exp \left[i \oint \cA^\theta_{I} |\dot x| ds\right]  \\
&\quad{}+ e^{-i\pi \cos\theta} \Tr \cP \exp \left[i \oint \left(\cA^\theta_{I+1}+\frac{1}{2}\cos\theta\right) |\dot x| ds\right].
\eal

To proceed with the deformation, we define as before the matrix
\beq
\label{cG-alpha}
\cG=
\left(
\begin{array}{c c c}
0&  \bar\alpha_{I-1} \tilde r_{I-1\, \dot 2} & 0 \\
\alpha_{I-1} \bar{\tilde r}_{I-1}^{\,\dot 2}& 0 &\alpha_I q_I^2 \\
0& \bar\alpha_I \bar{q}_{I2} & 0
\end{array}
\right).
\eeq
Crucially, this matrix is such that $i(\cQ_+)^2\cG=\partial_\varphi \cG-i[\cL_0^\theta,\cG]$. 
The expression in \eqref{cG-alpha} can 
then be used to construct a deformed 
superconnection
\beq
\label{cLalpha}
\cL_{\alpha,\bar\alpha}^\theta=\cL_0^\theta+i\cQ_+\cG+\cG^2\,.
\eeq
Under the action of $\cQ_+$
\begin{equation}
\label{Q+Ltheta2node}
\cQ_+\cL_{\alpha,\bar\alpha}^\theta=i(\cQ_+)^2\cG+\{\cQ_+\cG,\cG\}=\partial_\varphi\cG-i[\cL_0^\theta,\cG]+\{\cQ_+\cG,\cG\}\,.
\end{equation}
This is easily seen to correspond to the supercovariant derivative
\beq
\cD_\varphi\cG\equiv \partial_\varphi\cG-i\{\cL_{\alpha,\bar\alpha}^\theta,\cG]
\,,
\eeq
so that the hyperloop with superconnection $\cL_{\alpha,\bar\alpha}^\theta$ transforms as a total derivative 
under $\cQ_+$. 

To analyse the behavior under the action of $\cQ_-$ it is useful to split $\cG=G+\bar G$ with 
\begin{equation}
\label{GbarG}
G=\left(
\begin{array}{c c c}
0& 0 & 0 \\
\alpha_{I-1} \bar{\tilde r}_{I-1}^{\,\dot 2}& 0 &\alpha_I q_I^2 \\
0& 0 & 0
\end{array}
\right) \quad \text{and} \quad
\bar G=
\left(
\begin{array}{ccc}
0&  \bar\alpha_{I-1} \tilde r_{I-1\, \dot 2} & 0 \\
0 & 0 & 0 \\
0& \bar\alpha_I \bar{q}_{I2} & 0
\end{array}
\right)\,,
\end{equation}
which satisfy $\cQ_-G=\cQ_+G$ and $\cQ_-\bar G=-\cQ_+\bar G$. One has then
\bal
\label{Q-L}
\cQ_-\cL_{\alpha,\bar\alpha}^\theta
&=i(\cQ_+)^2(-G +\bar G)+\{\cQ_+(G -\bar G),\cG\}\cr
&=\partial_\varphi(-G+\bar G)-i[\cL^\theta_0,-G+\bar G]+\{\cQ_+(G -\bar G),\cG\}
\\
&=\cD_\varphi(-G+\bar G)
+\{\cQ_+(G -\bar G),\cG\}-\{\cQ_+\cG,-G+\bar G\}+i[\cG^2,-G+\bar G]\,,
\eal
where we attempted to write the last line as a covariant derivative with a 
$\cL^\theta_{\alpha,\bar\alpha}$ superconnection. For the loop to be invariant under $\cQ_-$, the 
three extra (anti)commutators should vanish and while there are some cancellations among them, 
the remaining terms are eliminated when $G$ and $\bar G$ are nilpotent of degree 2 as in (\ref{GbarG}).
This implies that for generic $\alpha$ and $\bar\alpha$ these 
Wilson loops are $1/8$ BPS, preserving exactly the same supercharges as $\cL_0^\theta$. 

Next we turn to the hyperloops with the quiver diagram in Figure~\ref{fig:1/4-quiver2}, but again 
we take multiplicity equal to one in the central node. In this case\footnote{Again, we always use the same symbols $\cL^\theta_0$ and $\cG$ and refer to explicit equations to avoid confusion.}
\beq
\label{Lbos3node-2}
\cL^\theta_0=\begin{pmatrix}
\cA_{I-1}^\theta & 0 & 0 \\
0 & \cA_I^\theta+ \frac{1}{2} & 0 \\
0& 0& \cA_{I+1}^\theta +\frac{1-\cos\theta}{2}
\end{pmatrix}\,.
\eeq
Note that now the connection in the bottom right corner is shifted by $(1-\cos\theta)/2$, 
though the corresponding node in the diagram is not squiggly.
This does not exactly match the shifts, which are no longer all equal to 1/2. 
However, we retain the same notation in the figure to indicate the allowed couplings to matter fields that 
preserve the supercharges. These are now organized in the matrix 
\beq
\label{G-3node-beta}
\cG=
\left(
\begin{array}{ccc}
0& \beta_{I-1}  \tilde r_{I-1\, \dot 1} & 0 \\
\bar\beta_{I-1} \bar{\tilde r}_{I-1}^{\,\dot 1}& 0 & \bar\beta_{I}  q_I^1\\
0& \beta_I \bar{q}_{I1} & 0
\end{array}
\right)\,,
\eeq
and the construction proceeds exactly as before to produce another family of $1/8$ BPS loops.

To construct $1/16$ BPS loops we again gauge transform the connection in \eqref{Lbos3node-2} 
to bring it to the form in \eqref{Lbos3node}. This results in the extra phases in the corresponding $\cG$,
\beq
\label{gauge-transformed}
\cG\to
\left(
\begin{array}{ccc}
0& \beta_{I-1} e^{i\varphi} \tilde r_{I-1\, \dot 1} & 0 \\
\bar\beta_{I-1} e^{-i\varphi} \bar{\tilde r}_{I-1}^{\,\dot 1}& 0 & \bar\beta_I e^{-i\varphi\cos\theta} q_I^1 \\
0& \beta_I e^{i\varphi\cos\theta} \bar{q}_{I1}  & 0
\end{array}
\right)\,.
\eeq
We can now add this expression to the $\cG$ in \eqref{cG-alpha} to get
\beq
\label{genericG}
\cG=
\left(
\begin{array}{c c c}
0&  \bar\alpha_{I-1} \tilde r_{I-1\, \dot 2}+ \beta_{I-1} e^{i\varphi} \tilde r_{I-1\, \dot 1} & 0 \\
\alpha_{I-1} \bar{\tilde r}_{I-1}^{\,\dot 2}+\bar\beta_{I-1} e^{-i\varphi} \bar{\tilde r}_{I-1}^{\,\dot 1}& 0 &
\alpha_I q_I^2+\bar\beta_I e^{-i\varphi\cos\theta} q_I^1 \\
0& \bar\alpha_I \bar{q}_{I2}+\beta_I e^{i\varphi\cos\theta} \bar{q}_{I1} & 0
\end{array}
\right).
\eeq
This is similar to the sum of the two matrices in \eqref{16GbarG}, but there is an obstruction to 
using this $\cG$ to construct a deformation, as the combination $\alpha_I q_I^2+\bar\beta_I e^{-i\varphi\cos\theta} q_I^1$ is not periodic. 
This is worse than the non-periodicity in \eqref{gauge-transformed}, as it cannot be fixed by a 
gauge transformation. The only way to overcome this is to set either $\alpha_I=\bar\alpha_I=0$ or 
$\beta_I=\bar\beta_I=0$, giving two branches of $1/16$ BPS Wilson loops.

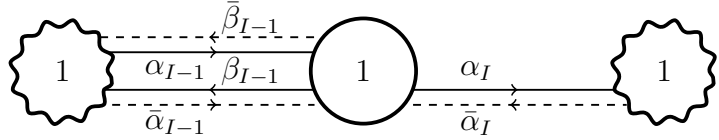
\begin{figure}[H]
\centering
\begin{tikzpicture}
\draw[decoration={snake,amplitude = .5mm,segment length=3.46mm},decorate, line width=.5mm] (2,1.9) circle (6.5mm);
\draw[line width=.5mm] (6,2) circle (7mm);
\draw[decoration={snake,amplitude = .5mm,segment length=3.46mm},decorate, line width=.5mm] (10,1.9) circle (6.5mm);
\draw[line width=.25mm, ->] (2.62,2.25)--(4.05,2.25);
\draw[line width=.25mm] (4,2.25)--(5.35,2.25);
\draw[line width=.25mm,dashed] (2.5,2.45)--(4,2.45);
\draw[line width=.25mm,dashed,<-] (3.95,2.45)--(5.45,2.45);
\draw[line width=.25mm] (2.55,1.75)--(4,1.75);
\draw[line width=.25mm,<-] (3.95,1.75)--(5.35,1.75);
\draw[line width=.25mm, dashed, ->] (2.66,1.55)--(4.05,1.55);
\draw[line width=.25mm,dashed] (4,1.55)--(5.35,1.55);
\draw[line width=.25mm, ->] (6.66,1.75)--(8.05,1.75);
\draw[line width=.25mm] (8,1.75)--(9.35,1.75);
\draw[line width=.25mm,dashed, ] (6.66,1.55)--(8,1.55);
\draw[line width=.25mm,dashed, <-] (7.95,1.55)--(9.5,1.55);
\draw (2,2) node  []  {$1$};
\draw (6,2) node  []  {$1$};
\draw (10,2) node  []  {$1$};
\draw (3.5,2) node  []  {$\alpha_{I-1}$};
\draw (3.5,1.3) node  []  {$\bar{\alpha}_{I-1}$};
\draw (4.5,2) node  []  {$\beta_{I-1}$};
\draw (4.5,2.7) node  []  {$\bar{\beta}_{I-1}$};
\draw (7.5,2) node  []  {${\alpha}_{I}$};
\draw (7.5,1.3) node  []  {$\bar\alpha_{I}$};
\end{tikzpicture}
\caption{A quiver diagram for a $1/16$ BPS Wilson loop.}
\label{fig:1/16-quiver-1}
\end{figure}

For $\beta_I=\bar\beta_I=0$ we can represent it by the quiver in Figure~\ref{fig:1/16-quiver-1}.

\begin{figure}[H]
\centering
\begin{tikzpicture}
\draw[decoration={snake,amplitude = .5mm,segment length=3.46mm},decorate, line width=.5mm] (6,1.9) circle (6.5mm);
\draw[line width=.5mm] (2,2) circle (7mm);
\draw[line width=.5mm] (10,2) circle (7mm);
\draw[line width=.25mm, ->] (2.65,2.25)--(4.05,2.25);
\draw[line width=.25mm] (4,2.25)--(5.45,2.25);
\draw[line width=.25mm,dashed] (2.55,2.45)--(4,2.45);
\draw[line width=.25mm,dashed,<-] (3.95,2.45)--(5.45,2.45);
\draw[line width=.25mm] (2.65,1.75)--(4,1.75);
\draw[line width=.25mm,<-] (3.95,1.75)--(5.35,1.75);
\draw[line width=.25mm, dashed, ->] (2.55,1.55)--(4.05,1.55);
\draw[line width=.25mm,dashed] (4,1.55)--(5.45,1.55);
\draw[line width=.25mm] (6.6,2.25)--(8,2.25);
\draw[line width=.25mm,<-] (7.95,2.25)--(9.35,2.25);
\draw[line width=.25mm, dashed,->] (6.55,2.45)--(8.05,2.45);
\draw[line width=.25mm,dashed ] (8,2.45)--(9.45,2.45);
\draw (2,2) node  []  {$1$};
\draw (6,2) node  []  {$1$};
\draw (10,2) node  []  {$1$};
\draw (3.5,2) node  []  {$\alpha_{I-1}$};
\draw (3.5,1.3) node  []  {$\bar{\alpha}_{I-1}$};
\draw (4.5,2) node  []  {$\beta_{I-1}$};
\draw (4.5,2.7) node  []  {$\bar{\beta}_{I-1}$};
\draw (8.5,2) node  []  {${\beta}_{I}$};
\draw (8.5,2.7) node  []  {$\bar\beta_{I}$};
\end{tikzpicture}
\caption{The other class of $1/16$ BPS loops.}
\label{fig:1/16-quiver-2}
\end{figure}
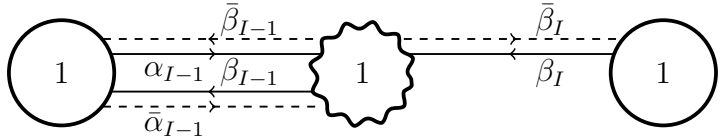

In the case of $\alpha_I=\bar\alpha_I=0$, it is a bit nicer to employ the gauge in 
\eqref{Lbos3node-2} to avoid the awkward phase multiplying $q_I^1$. The diagram is 
in Figure~\ref{fig:1/16-quiver-2}.

\subsection{A closer look at the hyperloops}
\label{sec:closer}

Let us examine in detail the hyperloops with superconnection $\cL_{\alpha,\bar\alpha}$ in \eqref{cLalpha}.
To do so we need the explicit expressions of $\cQ_+$ acting on the scalar fields $q^a_I$ and $\tilde r_{I-1 \, \dot a}$. 
For the latter, 
we get back components of $\tilde\psi^a_{I-1}$ as
(the $\pm$ on the $\tilde\psi$ represent spinor indices)
\beq
\cQ_+\tilde{r}_{I-1\,\dot{1}} = -\tilde\psi_{I-1,+}^1\,,
\quad
\cQ_+\bar{\tilde{r}}^{\dot 1}_{I-1}=\bar{\tilde{\psi}}_{I-1\,1,-}\,,\quad
\cQ_+\tilde{r}_{I-1\,\dot{2}} =\tilde\psi_{I-1,-}^2\,,
\quad
\cQ_+\bar{\tilde{r}}^{\dot 2}_{I-1}=-\bar{\tilde{\psi}}_{I-1\,2,+}\,.
\eeq
For the action on the $q^a_I$ fields it is convenient to define rotated fermions $\rho_{I  \dot a}$ via
\bea
\label{rho}
\rho_{I\dot{1}}\equiv \cos\frac{\theta}{2}~\psi_{I\dot 1}+ e^{-i\varphi}\sin\frac{\theta}{2}~\psi_{I\dot 2} \,,
\qquad
 \rho_{I\dot 2}\equiv\cos\frac{\theta}{2}~\psi_{I\dot 2}- e^{i\varphi} \sin\frac{\theta}{2}~\psi_{I\dot 1}\,,
\eea
such that 
\beq
\cQ_+q_I^1=-\rho_{I\dot{1},-}\,,
\quad
\cQ_+q_I^2=\rho_{I\dot{2},+}\,,
\quad
\cQ_+\bar{q}_{I1}=\bar\rho_{I,+}^{\dot 1}\,,
\quad
\cQ_+\bar q_{I2}=-\bar\rho_{I,-}^{\dot 2}\,.
\end{equation}
Hence we find
\beq
\label{closerlookL}
\cL^\theta_{\alpha,\bar\alpha}
=
\cL_0^\theta
+\begin{pmatrix}  
\bar\alpha_{I-1}\alpha_{I-1} \tilde r_{I-1 \,\dot 2}\bar{\tilde r}^{\,\dot 2}_{I-1} & 
i\bar\alpha_{I-1}\tilde\psi_{I-1,-}^2  &  
\bar\alpha_{I-1} \alpha_I \tilde r_{I-1 \,\dot 2}q^2_I \\
-i\alpha_{I-1}\bar{\tilde{\psi}}_{I-1\,2,+}& 
\alpha_{I-1}\bar\alpha_{I-1}\bar{\tilde r}^{\,\dot 2}_{I-1}\tilde r_{I-1\,\dot 2} + \alpha_I \bar\alpha_I  q^2_I \bar q_{I2}&
i\alpha_I\rho_{I\, \dot2,+}
\\
\bar\alpha_I \alpha_{I-1} \bar q_{I 2}\bar{\tilde r}_{I-1}^{\,\dot 2} & 
-i\bar\alpha_I\bar{\rho}^{\dot 2}_{I,-}&
\bar\alpha_I\alpha_I  \bar q_{I2}q^2_I
\end{pmatrix}.
\eeq

An interesting question is whether there are any points of enhanced supersymmetry along our 
moduli space. A simple guide is to look for points of enhanced bosonic symmetry that does not 
commute with the preserved supercharges, so either $SU(2)_L$ or $SU(2)_R$.

Looking to impose the $SU(2)_R$ symmetry, recall that it acts on the dotted indices of $\tilde r$ and $\tilde\rho$. Examining \eqref{closerlookL}, we immediately see that we should impose $\alpha_I=\bar\alpha_I=0$ to eliminate the off-diagonal entries where dotted indices appear. This has the effect of eliminating the entire third row and column of the supermatrix. The corresponding diagonal entry in $\cL^\theta_0$, which is $\cA^\theta_{I+1}$, is also clearly not $SU(2)_R$ symmetric, as can be seen from \eqref{AIpm1}. So we should remove it from $\cL^\theta_0$ as well and focus on a loop coupling to only two nodes.
Of course, it is possible to generalize this to any even number of nodes, but with only couplings between pairs.

Focusing then on the upper-left $2\times2$ block of \eqref{closerlookL}, there are still 
explicit dotted indices in the diagonal parts, but 
those appear also in $\cL^\theta_0$. Let us write them together, say for the $I$-th node:
\beq
\label{genericM3node}
\cA^\theta_I+\alpha_{I-1}\bar\alpha_{I-1}\bar{\tilde r}^{\,\dot 2}_{I-1}\tilde r_{I-1\,\dot 2} + \alpha_I \bar\alpha_I  q^2_I \bar q_{I2}
=A_{\varphi,I}
-\frac{i}{k}
\bar{\tilde r}_{I-1}^{\,\dot a}
\widetilde M_{\dot a}{}^{\dot b}
\tilde r_{I-1\,\dot b}
-\frac{i}{k}
q_{I}^a
M_a{}^b
\bar q_{I\,b}\,,
\eeq
with
\begin{equation}
\label{scalarcouplings}
\widetilde M_{\dot a}{}^{\dot b}=\begin{pmatrix}
1\; &\;
0 \\
0 \; &\;  ik\alpha_{I-1}\bar\alpha_{I-1}-1
\end{pmatrix} \quad\text{and}\quad 
M_a{}^b =\begin{pmatrix}
1 \; & \; 0  \\
0 \; & \;  ik \bar\alpha_I \alpha_I-1
\end{pmatrix}\,.
\end{equation}
We chose to write the matrix $\widetilde M$ in the basis of the twisted $\tilde r$, but regardless of 
the basis, in order to preserve $SU(2)_R$, it has to be proportional to the identity, so 
$ik\alpha_{I-1}\bar\alpha_{I-1}=2$.

The exact same structure follows for the top left entry, which is now also symmetric. Recalling the 
$\bC^*$ symmetry, we can further fix $\alpha_{I-1}=\bar\alpha_{I-1}=\sqrt{2i/k}$ and the full form of 
the connection becomes
\beq
\label{SU(2)R1}
\cL^\theta
=
\begin{pmatrix}  
A_{\varphi,I-1}+\tilde q_{I-1 \,\dot a}\bar{\tilde q}^{\,\dot a}_{I-1}
+\bar q_{I1}q^1-\bar q_{I2}q^2+\frac{1}{2} & 
-\frac{1-i}{\sqrt k}\,\tilde\psi_{I-1,-}^2  \\
\frac{1-i}{\sqrt k}\,\bar{\tilde{\psi}}_{I-1\,2,+}& 
A_{I,\varphi}+\bar{\tilde q}^{\,\dot a}_{I-1}\tilde q_{I-1 \,\dot a}
+q^1\bar q_{I1}-q^2\bar q_{I2}
\end{pmatrix}.
\eeq
Note that the $\theta$-dependence completely dropped out of this expression, so it is within 
the class of 1/4 BPS operators presented in Section~\ref{sec:theta=0}, but with extra 
$SU(2)_R$ symmetry, so this is in fact the 1/2 BPS loop of \cite{Cooke:2015ila}, now adapted 
to the 3-sphere.

Imposing $SU(2)_L$ symmetry is similar, but the results are different. Looking for undotted 
indices in \eqref{closerlookL}, we see that now we should take $\alpha_{I-1}=\bar\alpha_{I-1}=0$. 
This eliminates the top line and left column from the matrix and, as before, we should remove 
the top left entry in $\cL_0^\theta$ as well.

We should again also examine the diagonal blocks, as we again have the expression in 
\eqref{genericM3node}.  Now $\widetilde M=\diag(1,-1)$ and we want $M$ to be proportional 
to the identity, so we set $ik\bar\alpha_I\alpha_1=2$, or $\alpha_I=\bar\alpha_I=\sqrt{2i/k}$. 
The resulting $2\times2$ connection is
\beq
\label{SU(2)L1}
\cL^\theta
=
\begin{pmatrix}  
A_{I,\varphi}+ \bar{\tilde r}_{I-1}^{\dot1}\tilde r_{I-1 \,\dot 1}-\bar{\tilde r}_{I-1}^{\dot2}\tilde r_{I-1 \,\dot 2}+q^a\bar q_{Ia}&
-\frac{1-i}{\sqrt k}\,\rho_{I\, \dot2,+}
\\
\frac{1-i}{\sqrt k}\,\bar{\rho}^{\dot 2}_{I,-}&
A_{I+1,\varphi}+ \tilde r_{I-1 \,\dot 1}\bar{\tilde r}_{I-1}^{\dot1}-\tilde r_{I-1 \,\dot 2}\bar{\tilde r}_{I-1}^{\dot2}+\bar q_{Ia}q^a
+\frac12
\end{pmatrix}.
\eeq
The fermions in the off-diagonal entries are defined in terms of the original fields in \eqref{rho} and 
the scalar bilinears are
\beq
\bar{\tilde r}_{I-1}^{\dot1}\tilde r_{I-1 \,\dot 1}-\bar{\tilde r}_{I-1}^{\dot2}\tilde r_{I-1 \,\dot 2}
=
\begin{pmatrix}
\bar{\tilde q}_{I-1}^{\,\dot1}&\bar{\tilde q}_{I-1}^{\,\dot2}
\end{pmatrix}
\begin{pmatrix}
\cos\theta & e^{-i\varphi}\sin\theta  \\
e^{i\varphi}\sin\theta & -\cos\theta 
\end{pmatrix}
\begin{pmatrix}
\tilde q_{I-1\,\dot1}\\\tilde q_{I-1\,\dot2}
\end{pmatrix},
\eeq
We recognize the same structure as the ``fermionic latitude'' of \cite{Cardinali:2012ru}, so this is 
its generalization to arbitrary $\cN=4$ Chern-Simons-matter theories. 
For $\theta=0$ it becomes another class of $1/2$ BPS loop of \cite{Cooke:2015ila}.

In addition to the supercharges $\cQ_\pm$, this loop is also invariant under 
the ones obtained by swapping undotted indices,
\beq
\label{supercharges2}
\cos\frac{\theta}{2}\,Q^{\dot 12}_{\bar l}+\sin\frac{\theta}{2}\,Q^{\dot22}_{\bar r}\,,
\qquad
\cos\frac{\theta}{2}\,Q^{\dot 21}_{l}-\sin\frac{\theta}{2}\,Q^{\dot11}_{r}\,.
\eeq

The examples thus far were for the grading in Figure~\ref{fig:1/4-quiver}. The story for the second 
grading is analogous, with the $\beta$ parameters coupling to the remaining fields. In particular,
 the matrices $\widetilde M$ and $M$ in \eqref{scalarcouplings} would have the upper left corner shifted 
by bilinears of $\beta$. To preserve $SU(2)_R$ we now set $\beta_I=\bar\beta_I=0$, and again 
have to focus on a superconnection involving only the $I-1$ and $I$ nodes with $M$ as 
before and $\widetilde M=\diag(-1,-1)$. This is the second $SU(2)_R$ invariant 1/2 BPS 
Wilson loop described in \cite{Cooke:2015ila}. Likewise, the $SU(2)_L$ invariant hyperloop on this branch is another version of 
the ``fermionic latitude'' loops described in \cite{Cardinali:2012ru} (in the notation of that paper it has 
$l=-1$).

The analysis here relies on a bosonic symmetry to indicate enhanced supersymmetry. In 
principle there could be further points with accidental or more subtle supersymmetry enhancement. 
We leave the study of that to the future.


\subsection{Further examples}

So far we focused on hyperloops involving three nodes of a long quiver. The simplest generalization arising already in that case is taking multiple copies of the connections, i.e. $p_J>1$ in 
Figures~\ref{fig:1/4-quiver} and~\ref{fig:1/4-quiver2}, forming larger $\cL_0^\theta$ connections and 
involving more $\alpha$ and/or $\beta$ parameters in the appropriate matrices $\cG$. This 
mirrors the examples in Section~\ref{sec:theta=0}, where the central node had multiplicity 2.

In the case of a linear quiver one may worry about the final nodes which couple either to hypermultiplets 
or to twisted hypermultiplets, although both kinds of fields appear in \eqref{1/8} and \eqref{AIpm1}. This turns 
out not to be a problem, and the construction proceeds as before with the missing moment 
maps removed. For example, suppose that there is no hypermultiplet to the left of the $I-1$ node 
in the quiver of the underlying theory. When constructing the hyperloop as in \eqref{Lbos3node}, 
this would mean that $\cA_{I-1}^\theta$ would loose the $\mu$ contribution and become
\beq
A_{\varphi,I-1}-\frac{i}{k}(\tilde{r}_{I-1\,\dot1}\bar{\tilde{r}}_{I-1}^{\dot1}-\tilde{r}_{I-1\,\dot2}\bar{\tilde{r}}_{I-1}^{\dot2})\,,
\eeq
but the construction of the hyperloop would follow exactly as before.
Likewise if the underlying theory had no twisted hypermultiplet to the right of the node 
$I+1$, then $\cA_{I+1}^\theta$ would become
\beq
A_{\varphi,I+1}
-\frac{i}{k}({{\mu_{I+1}}_1}^{1}-{{\mu_{I+1}}_2}^{2})\,,
\eeq
and again the construction would follow as before.

Of course, if we have only a 2-node quiver and no twisted hypermultiplets, there would not 
be a way to $\theta$-deform their couplings. In that case there would only be the analog 
deformation of $\mu$, which is completely parallel to the constructions based on $\tilde\mu$.

\begin{figure}[H]
\centering
\begin{tikzpicture}
\draw[decoration={snake,amplitude = .5mm,segment length=3.46mm},decorate, line width=.5mm] (2,1.9) circle (6.5mm);
\draw[line width=.5mm] (6,2) circle (7mm);
\draw[decoration={snake,amplitude = .5mm,segment length=3.46mm},decorate, line width=.5mm] (10,1.9) circle (6.5mm);
\draw[line width=.5mm] (14,2) circle (7mm);
\draw[line width=.25mm, dashed, ->] (2.66,1.6)--(4.05,1.6);
\draw[line width=.25mm,dashed] (4,1.6)--(5.35,1.6);
\draw[line width=.25mm] (2.6,1.8)--(4,1.8);
\draw[line width=.25mm,<-] (3.95,1.8)--(5.35,1.8);
\draw[line width=.25mm, ->] (6.66,1.8)--(8.05,1.8);
\draw[line width=.25mm] (8,1.8)--(9.35,1.8);
\draw[line width=.25mm,dashed, ] (6.66,1.6)--(8,1.6);
\draw[line width=.25mm,dashed, <-] (7.95,1.6)--(9.5,1.6);
\draw[line width=.25mm, dashed, ->] (10.66,1.6)--(12.05,1.6);
\draw[line width=.25mm,dashed] (12,1.6)--(13.35,1.6);
\draw[line width=.25mm] (10.6,1.8)--(12,1.8);
\draw[line width=.25mm,<-] (11.95,1.8)--(13.35,1.8);
\draw (2,2) node  []  {$p_{I-1}$};
\draw (6,2) node  []  {$p_{I}$};
\draw (10,2) node  []  {$p_{I+1}$};
\draw (14,2) node  []  {$p_{I+2}$};
\draw (4.05,2) node  []  {$\alpha_{I-1}$};
\draw (4.05,1.3) node  []  {$\bar{\alpha}_{I-1}$};
\draw (8,2) node  []  {${\alpha}_{I}$};
\draw (8,1.3) node  []  {$\bar\alpha_{I}$};
\draw (12.05,2) node  []  {$\alpha_{I+1}$};
\draw (12.05,1.3) node  []  {$\bar{\alpha}_{I+1}$};
\end{tikzpicture}
\caption{A quiver diagram for a 4-node 1/8 BPS hyperloop.}
\label{fig:4-nodes}
\end{figure}
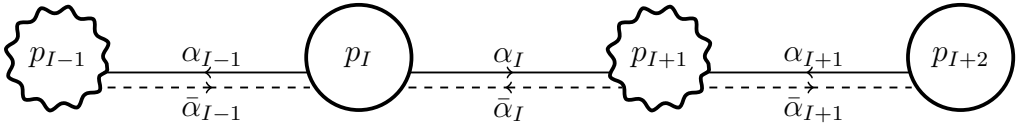

Next we examine what  happens for longer quivers and for circular quivers like ABJ(M) theory. 
Consider first the hyperloop coupling to 4 nodes as in Figure~\ref{fig:4-nodes}.
The story proceeds exactly as before, except that we should remember that the shifts are relative 
to neighboring nodes, so in this case (for all $p_J=1$) the starting point is
\beq
\cL_0^\theta=\begin{pmatrix}
\cA_{I-1}+\frac{1}{2}&0&0&0\\
0&\ \cA_{I}\ &0&0\\
0&0&\cA_{I+1}+\frac{1}{2}\cos\theta&0\\
0&0&0&\cA_{I+2}+\frac{\cos\theta-1}{2}
\end{pmatrix}.
\eeq
It should be clear how to deform this loop by adding couplings to the fermions and also construct the 
loop corresponding the quiver with the second possible grading.

In the case of circular quivers, turning on $\theta\neq0$ poses a challenge. In the simplest case of ABJ(M) we have 
two nodes, so $L=2$. Then in our figures and expressions for $\cL_0$, etc. we can take $I=2$ and identify the nodes
 $I-1=1$ and $I+1=3$. If we consider a hyperloop coupling
only to one edge in the quiver---so the quiver for the hyperloop is a linear 2-node quiver---there is no problem. If 
we want to couple to matter from both nodes, we face the fact that while we identify nodes 1 and 3, the shift 
of $\cA_1$ is $\frac12$ and that of $\cA_3$ it should be $\frac1{2}\cos\theta$, see \eqref{Lbos3node}.

The solution to this problem was already anticipated in \cite{drukker2020bps} (in other contexts where the shifts were not 
$1/2$) and it amounts to taking a cover of the original gauge theory quiver. So we can couple the Wilson 
loop to both edges of the original quiver as long as we consider a 3-node hyperloop with $\cL_0^\theta$ given by
\beq
\label{ABJ(M)3}
\cL^\theta_0=\begin{pmatrix}
\cA_{1}+\frac{1}{2}&0&0\\
0&\ \cA_{2}\ &0\\
0&0&\cA_{1}+\frac{1}{2}\cos\theta
\end{pmatrix}.
\eeq
Here $\cA_1$ appears twice with different shifts. One can continue further with another copy of $\cA_2$ with 
a shift of $(\cos\theta-1)/2$, and so on.

For $\theta\neq0$ the connection in \eqref{ABJ(M)3} has $(\bC^*)^3$ symmetry (with one copy acting trivially), 
so the moduli space of $1/4$ BPS loops according to \eqref{open-quiver-moduli} is
\beq
\bC^{2}\times\bC^{2}/\!\!/\bC^*\times\bC^*\simeq\bC^2\,.
\eeq
The $\theta=0$ case has enhanced symmetry 
$(\bC^*)^3\to GL(2,\bC)\times \bC^*$. The moduli space 
according to \eqref{ABJ(M)-moduli} is now four-dimensional
\beq
\bC^{8}/\!\!/GL(2,\bC)\,.
\eeq
In addition to the very special case of $\theta=0$, for any rational $\cos\theta$, a hyperloop based 
on a long enough quiver will have also some enhanced symmetry.


\section{A matrix model proposal}
\label{sec:MM}

The construction of the hyperloops is based on a deformation of a bosonic loop and all the loops 
of fixed $\theta$ are cohomologically equivalent under the supercharge $\cQ_+$ used to define them. 
This means that any localization computation for any loop on the moduli space is immediately applicable 
to any one. The (tedious) proof, following \cite{Drukker:2009hy,Cooke:2015ila} requires expanding 
the exponentials and checking order-by-order that the difference between the different operators is 
$\cQ_+$-exact. We do not reproduce this computation here since it is essentially identical.

We propose now a matrix model that we hope captures the expectation value of our operators. The matrix model partition function can be motivated by considering the usual ingredients due to the vector multiplets (hyperbolic sines) and hypermultiplets (hyperbolic cosines) at each node \cite{Kapustin:2009kz} and the proposal in \cite{Bianchi:2018bke} on how to introduce the $\theta$-deformation:
\bal
\label{ZMMproposal}
Z=& \prod_I\frac{1}{N_I!} \int \prod_{i=1}^{N_I}\frac{d\lambda_{Ii}}{2\pi}e^{i\frac{k_I}{4\pi} \lambda_{Ii}^2}
\prod_{i<j}^{N_I}4 \sinh \frac{\nu (\lambda_{Ii}-\lambda_{Ij})}{2} \sinh\frac{ \lambda_{Ii}-\lambda_{Ij}}{2 \nu}
 \cr
& \hskip 2.2cm \times
\left( \prod_{i=1}^{N_I}\prod_{j=1}^{N_{I+1}}
2 \cosh \frac{\nu^{(-1)^I} (\lambda_{Ii}-\lambda_{I+1,j})}{2} 
\right)^{-1} \,.
\eal
The explicit value of the parameter $\nu$ can be fixed by the comparison with a perturbative computation and turns out to be given, in our notation, by $\nu=\sqrt{\cos\theta}$ \cite{Bianchi:2018bke}. 
Note that twisted and untwisted hypermultiplets contribute differently, with $\nu$ either in the numerator or in the denominator of the argument of the hyperbolic cosines. The expectation value of the 
$\theta$-deformed hyperloops 
is given by inserting in the partition function above  $\sum_I \sum_{i=1}^{N_I} e^{\nu\lambda_{Ii}}$. 


\section{Conclusions}
\label{sec:conclusions}

This paper reorganized the space of known Wilson loop operators in $\cN=4$ Chern-Simons-matter 
theories in three dimensions, which we now call hyperloops, and generalized it considerably to include 
loops preserving 1, 2, 4 and 8 supercharges. Our findings clarify and elaborate the intricate structure 
of the supersymmetric line operators and their moduli spaces. The strategy, adapted from 
\cite{Drukker:2019bev, drukker2020bps}, is to start with a choice of diagonal (bosonic) superconnection, 
which is at the apex of a conical moduli space, and describes all other superconnections as deformations of it. 
The operators we have thus obtained are classified in terms of quiver diagrams encoding 
which gauge fields are involved and the couplings to the matter fields. This, together with 
a ``latitude'' parameter $\theta$, completes the set of data necessary for the classification.

More concretely, we need to choose some or all of the vector fields and the number of 
times they are represented in the Wilson loop. The next step is to choose a grading, which 
plays two roles: it indicates the constant shifts in some of the diagonal connections 
and it implies which half of the chiral fields we couple to, in order to get hyperloops with 
twice as many supercharges. The two possible gradings then also give the two branches 
of the moduli space. Finally, for each of the included chiral (antichiral) fields we 
have couplings $\alpha_I$, $\beta_I$ ($\bar\alpha_I$, $\bar\beta_I$) 
subject to a global gauge invariance, reducing the moduli space to a quotient of 
$\bC^p$ for some $p$. These moduli spaces are known as quiver varieties and generically are 
conical, as first observed in \cite{Drukker:2019bev} for the conifold in the 
case of the 1/6 BPS loops of the ABJ(M) theory.

The hyperloops with $\theta=0$ were previously found in \cite{drukker2020bps}, but our description is much more 
algorithmic and their moduli space was never studied in such detail.
Loops with $\theta\neq0$ were only studied in ABJ(M) theory and not in theories with $\cN=4$ supersymmetry. 
Furthermore, the constructions focused on the bosonic loops and on the analogs of 
the 1/4 BPS loops (which are 1/6 BPS in ABJ(M)) presented in Section~\ref{sec:closer}. The continuous 
family of hyperloops interpolating between those two cases and all the other directions in 
the moduli space have not been previously described.

As mentioned in Section~\ref{sec:bos}, the Gaiotto-Yin loop breaks 
$SU(2)_L\times SU(2)_R\to U(1)_L\times U(1)_R$. 
The parameter $\theta$ further breaks $U(1)_R$ by including extra 
couplings to the moment maps arising from the twisted hypermultiplets. All the examples that 
are presented in the preceding sections have analogs with the roles of hypermultiplets and 
twisted hypermultiplets reversed.

There are many possible directions that can be pursued from here. Among the most 
obvious ones is to attempt the complete exploration of the full moduli space of line 
operators in three-dimensional Chern-Simons-matter theories. In addition to the tried 
and tested approach of making ans\"atze and restricting them to be BPS, we can try to 
extend the point of view introduced in \cite{Drukker:2019bev} and employed here, 
of constructing the loops as deformations of previously identified ones. One should 
first verify whether there are further line operators involving only single nodes. Then 
whether there are any further deformations of them with more complicated forms than 
considered here. Finally, one should examine other points along 
the moduli space to see whether there are other branches that may intersect those 
points, but not pass through the origin we employ here.

Still, these explorations cannot answer the question of what is the full space of BPS line 
operators, which would require new tools to address. Moreover, there are other types 
of line operators, known as vortex loops \cite{Drukker:2008jm, Drukker:2012sr, Kapustin:2012iw}. 
In some cases they are known to 
be dual under mirror symmetry to Wilson loops \cite{Assel:2015oxa}, so it should be exciting to 
understand their moduli spaces as well.

In Section~\ref{sec:closer} we looked at some special examples of these loops which 
have enhanced supersymmetry. It is not clear that the ones identified there, which were all 
previously known, are the only points of enhanced supersymmetry on the moduli space.

Also in this spirit of discovering new hyperloops would be the construction of operators 
supported along generic curves on a $S^2\subset S^3$, following the four-dimensional example of 
\cite{Drukker:2007dw,Drukker:2007yx,Drukker:2007qr} or the ABJ(M) analog of \cite{Cardinali:2012ru}.

Another question worth asking is what happens to these operators at the quantum level and 
whether the classical moduli spaces described here receive corrections. The analysis of 
\cite{Agmon:2020pde} suggests that moduli spaces are natural for line operators in three dimensions, but it does not predict their dimensions, as we found here, nor that their classical structure is not subject 
to quantum corrections. The heroic 3-loop calculation of \cite{Bianchi:2016vvm}\footnote{See also \cite{Griguolo:2015swa} for a previous attempt limited to a two-loop computation.} 
suggests that the degeneracy among pairs of 1/2 BPS Wilson loops of \cite{Cooke:2015ila} 
may sometimes get lifted. Such a perturbative analysis would also be useful to test the matrix model 
proposal we put forward in Section~\ref{sec:MM}.

Yet another angle is to study these moduli spaces as defect conformal manifolds in the context 
of defect CFT. Explicit analysis of this type for line operators in four-dimensional $\cN=4$ 
theory include \cite{Cooke:2017qgm, Giombi:2017cqn, Liendo:2018ukf} and in three dimensions \cite{Bianchi:2020hsz}.

Wilson loops are also interesting in the holographic context, for they provide a rich dictionary 
between gauge theory and string theory objects. It would then be interesting to understand the holographic realization of the operators 
constructed here. Very little has been done in this direction since the original proposal for the 
holographic dual in \cite{Drukker:2008zx, Chen:2008bp,Rey:2008bh}. Proposals for the 
holographic duals of $1/2$ BPS loops in some $\cN=4$ theories were put forward in 
\cite{Lietti:2017gtc} and a first examination of a possible moduli space of 1/6 BPS loops in ABJ(M) theory  
was done in \cite{Correa:2019rdk}.


\subsection*{Acknowledgements}

We would like to thank Luca Griguolo, Luigi Guerrini and Domenico Seminara for discussions. ND is supported by the Science Technology \& Facilities Council under the grants  ST/T000759/1 and ST/P000258/1.
MT acknowledges the support of the Conselho Nacional de Desenvolvimento Cientifico e Tecnologico (CNPq).
DT is supported in part by the INFN grant {\it Gauge and String Theory (GAST)} and would like to thank FAPESP's partial support through the grants 2015/17885-0,  2016/01343-7 and 2017/50435-4. 


\appendix

\section{The transformations for $\cN=4$ on $S^3$}
\label{app:susytransformations}

The supersymmetry transformations of the $\cN=4$ Chern-Simons-matter theory on the $S^3$ can be seen to be given by 
\bal
\label{SUSY2}
\delta A_{\mu\,I}&=\frac{i}{k}\xi_{a\dot b}\gamma_\mu(j_I^{a\dot b}-\tilde\jmath_I^{\,\dot ba})\,,
\hskip-3cm\\
\delta q_I^a&=\xi^{a\dot b}\psi_{I\,\dot b}\,,
\qquad&
\delta\bar q_{I\,a}&=\xi_{a\dot b}\bar\psi_I^{\dot b}\,,\\
\delta \tilde q_{I-1\,\dot b}&=-\xi_{a\dot b}\tilde\psi_{I-1}^{a}\,,
\qquad&
\delta\bar{\tilde q}_{I-1}^{\,\dot b}&=-\xi^{a\dot b}\bar{\tilde\psi}_{I-1\,a}\,,
\hskip8cm\\
\delta\psi_{I\,\dot a}&=i\gamma^\mu\xi_{b\dot a}D_\mu q_I^b
+i\zeta_{b\dot a} q_I^b
-\frac{i}{k}\xi_{b\dot a}(\nu_I q_I^{b}-q_I^b\nu_{I+1})
+\frac{2i}{k}\xi_{b\dot c}\left(
\tilde\mu_{I}{}_{\ \dot a}^{\dot c}q_I^b
-q_I^b\tilde\mu_{I+1}{}_{\dot a}^{\ \dot c}\right),
\hskip-10cm
\\
\delta\bar\psi_{I}^{\dot a}&=i\gamma^\mu\xi^{b\dot a}D_\mu \bar q_{I\,b}
+i\zeta^{b\dot a}\bar q_{I\,b}
-\frac{i}{k}\xi^{b\dot a}(\bar q_{I\,b}\nu_I -\nu_{I+1}\bar q_{I\,b})
+\frac{2i}{k}\xi^{b\dot c}\left(
\bar q_{I\,b\,}\tilde\mu_{I}{}_{\ \dot c}^{\dot a}
-\tilde\mu_{I+1}{}_{\dot c}^{\ \dot a}\bar q_{I\,b}\right),
\hskip-11cm
\\
\delta\tilde\psi_{I-1}^{a}&=-i\gamma^\mu\xi^{a\dot b}D_\mu \tilde q_{I-1\,\dot b}
-i\zeta^{a\dot b}\tilde q_{I-1\,\dot b}
+\frac{i}{k}\xi^{a\dot b}(\tilde q_{I-1\,\dot b}\tilde\nu_{I} -\tilde\nu_{I-1}\tilde q_{I-1\,\dot b})
\hskip-10cm
\\&\quad
-\frac{2i}{k}\xi^{b\dot c}\left(
\tilde q_{I-1\,\dot c\,}\mu_{I}{}_{\ b}^{a}
-\mu_{I-1}{}_{b}^{\ a}\tilde q_{I-1\,\dot c}\right),
\hskip-11cm
\\
\delta\bar{\tilde\psi}_{I-1\,a}&=-i\gamma^\mu\xi_{a\dot b}D_\mu \bar{\tilde q}_{I-1}^{\,\dot b}
-i\zeta_{a\dot b}\bar{\tilde q}_{I-1}^{\,\dot b}+\frac{i}{k}\xi_{a\dot b}(\tilde\nu_I \bar{\tilde q}_{I-1}^{\,\dot b}-\bar{\tilde q}_{I-1}^{\,\dot b}\tilde\nu_{I-1})
\hskip-10cm
\\&\quad
-\frac{2i}{k}\xi_{b\dot c}\left(
\mu_{I}{}_{\ a}^{b}\bar{\tilde q}_{I-1}^{\,\dot c}
-\bar{\tilde q}_{I-1}^{\,\dot c}\mu_{I-1}{}_{a}^{\ b}\right).
\hskip-11.5cm
\eal
where $\zeta_{a\dot b}=\frac{1}{3}\gamma^\mu \nabla_\mu \xi_{a\dot b}$. 
More specifically, from (\ref{xi}) one finds $\zeta^{l,\bar l}_{a\dot b}=\frac{i}{2}\xi^{l,\bar l}_{a\dot b}$ and $\zeta^{r,\bar r}_{a\dot b}=-\frac{i}{2}\xi^{r,\bar r}_{a\dot b}$. We work in Euclidean signature and take the gamma-matrices, $(\gamma^\mu)_\alpha^{\;\;\beta}$, to be given by the Pauli matrices. As usual, the spinor contractions are such that
\bea
\xi_1\xi_2\equiv \xi_1^\alpha \xi_{2,\alpha} =+\xi_ 2\xi_1\,,\qquad 
\xi_1 \gamma^\mu \xi_2\equiv 
\xi_1^\alpha (\gamma^\mu)_\alpha^{\;\;\beta} \xi_{2,\beta}
=-\xi_2 \gamma^\mu \xi_1\,,\qquad \alpha,\beta=\pm.
\eea
Then it follows that the Killing spinors on $S^3$ of \eqref{killingspinors} satisfy $\xi^{\bar l}\xi^l=\xi^l \xi^{\bar l}=1$ and $\xi^{\bar l} \gamma^\mu \xi^{l}=-\xi^{l} \gamma^\mu \xi^{\bar l}=\delta^\mu_\varphi$, and similarly for the contractions involving $\xi^r$ and $\xi^{\bar r}$.

The expressions in (\ref{SUSY2}) can be motivated by relating them to the transformations of the $\cN=2$ 
Chern-Simons-matter theory on $S^3$ written down in \cite{Hama:2010av,asano2012large}.  The off-shell supersymmetry transformation of the physical fields from the vector multiplets 
($A_\mu$, $\sigma$, $D$, $\lambda$, $\bar\lambda$) 
and chiral multiplets ($\phi$, $\psi$, $F$) in the $(\Box,\bar\Box)$ bifundamental and their conjugates 
are
\bal
\label{N=2off-shellsusy}
\delta A_{\mu\,I}&= -\frac{i}{2}(\bar\epsilon\gamma_\mu\lambda_I-\bar\lambda_I\gamma_\mu\epsilon)\,,\qquad
\delta\phi_I=\bar\epsilon\psi_I\,,\qquad
\delta\bar\phi_I=\epsilon\bar\psi_I\, ,\cr
\delta\psi_I&= i\gamma^\mu\epsilon D_\mu\phi_I + i\epsilon(\sigma_I\phi_I-\phi_I\sigma_{I+1})+\frac{i}{3}\gamma^\mu D_\mu\epsilon \phi_I + \bar\epsilon F_I\,,\cr
\delta\bar\psi_I&=i\gamma^\mu\bar\epsilon D_\mu\bar\phi_I -i(\sigma_{I+1}\bar\phi_I-\bar\phi_I\sigma_I)\bar\epsilon+\frac{i}{3}\bar\phi_I\gamma^\mu D_\mu\bar\epsilon+ \bar{F}_I\epsilon\,.
\eal
To match with $\cN=4$ theories we go on-shell, so use the actions%
\footnote{Note that for the $\cN=4$ theory the CS-levels are alternating, $k_I=(-1)^Ik$.}
\beq
S_{CS}^{(I)} = -\frac{k_I}{4\pi}  \int d^3x \sqrt{g} \Tr \left[ \frac{\epsilon^{\mu\nu\rho}}{\sqrt{g}}(A_{\mu\,I}\partial_\nu A_{\rho\,I}-\frac{2i}{3}A_{\mu\,I}A_{\nu\,I}A_{\rho\,I})-\bar\lambda_I\lambda_I+2 D_I \sigma_I\right],
\eeq
\bal
iS_\text{matter}^{(I)} = \int d^3x \sqrt{g} \Tr&\bigg[i D_\mu\bar\phi_I D^\mu\phi_I + \bar\psi_I\gamma^\mu D_\mu \psi_I + \frac{3i}{4l^2}\bar\phi_I\phi_I - \bar\psi_I(\sigma_I \psi_I - \psi_I\sigma_{I+1})\nonumber\\
&-\bar\psi_I(\lambda_I\phi_I-\phi_I \lambda_{I+1})+ \bar\phi_I(\bar\lambda_I\psi_I-\psi_I\bar\lambda_{I+1})-\bar\phi_I(D_I\phi_I -\phi_I D_{I+1})\\
&+i\bar\phi_I(\sigma_I\sigma_I\phi_I-2\sigma_I\phi_I\sigma_{I+1}+\phi_I\sigma_{I+1}\sigma_{I+1})+ i \bar{F}_I F_I\bigg]\,,
\eal
to integrate out the auxiliary fields $\lambda$, $\bar\lambda$, $\sigma$, $D$, $F$ and $\bar F$.

This yields the on-shell transformations
\begin{align}
\label{N=2on-shellsusy}
\delta A_{\mu\,I}&= -\frac{2\pi i}{k_I}\bigg[\bar\epsilon\gamma_\mu(\bar\phi_{I-1}\psi_{I-1}-\psi_I\bar\phi_I)-(\phi_{I}\bar\psi_{I}-\bar\psi_{I-1}\phi_{I-1})\gamma_\mu\epsilon\bigg]\,,\quad
\delta\phi_I=\bar\epsilon\psi_I\,,\quad
\delta\bar\phi_I=\epsilon\bar\psi_I\,,\cr
\delta\psi_I&= i\gamma^\mu\epsilon D_\mu\phi_I +  \frac{2\pi i}{k_I}\epsilon \bigg[(\bar\phi_{I-1}\phi_{I-1}-\phi_I\bar\phi_I)\phi_I-\phi_I(\phi_{I+1}\bar\phi_{I+1}-\bar\phi_{I}\phi_{I})\bigg]+\frac{i}{3}\gamma^\mu D_\mu\epsilon \phi_I \,,\nonumber\\
\delta\bar\psi_I&=i\gamma^\mu\bar\epsilon D_\mu\bar\phi_I -\frac{2\pi i}{k_I}\bigg[(\phi_{I+1}\bar\phi_{I+1}-\bar\phi_{I}\phi_{I})\bar\phi_I-\bar\phi_I(\bar\phi_{I-1}\phi_{I-1}-\phi_I\bar\phi_I)\bigg]\bar\epsilon+\frac{i}{3}\bar\phi_I\gamma^\mu D_\mu\bar\epsilon\,.
\end{align}

Now we use the chiral decomposition of $\cN=4$ hypermultiplets and twisted hypermultiplets in 
Figure~\ref{fig:N=4/2quiver}. The chiral fields in this representation are $q^2$ and $\tilde q_{\dot 1}$. 
Extending to the other fields in the multiplets we match \eqref{SUSY2} to \eqref{N=2on-shellsusy} with 
the replacements
\beq
\tilde{q}_{\dot{1}},\,q^2 \rightarrow \phi\,,\qquad 
\psi_{\dot{2}},\,-\tilde{\psi}^1\rightarrow \psi\,, \qquad
\bar{\tilde{q}}^{\dot{1}},\,\bar{q}_2 \rightarrow \bar\phi\,,\qquad
\bar\psi^{\dot{2}},\,-\bar{\tilde{\psi}}_1\rightarrow \bar\psi\,.
\eeq
where we also identified the supersymmetry parameters as 
$\xi^{1\dot{1}}\rightarrow\epsilon$ and $\xi^{2\dot{2}}\rightarrow\bar\epsilon$. A mismatch by $2\pi$ in 
the non-linear terms can be fixed by rescaling the fields.

The other fields transform in the conjugate $(\bar\Box,\Box)$ representation, and with the same choice of 
supersymmetry parameter identification they would be matched to $\cN=2$ fields in this representation 
according to
\beq
\bar{\tilde{q}}^{\dot{2}},\,\bar{q}_1 \rightarrow \phi\,,\qquad 
\bar\psi^{\dot{1}},\,-\bar{\tilde{\psi}}_2 \rightarrow \psi\,, \qquad
\tilde{q}_{\dot{2}},\,q^1 \rightarrow \bar\phi\,,\qquad
\psi_{\dot{1}},\,\,-\tilde{\psi}^2 \rightarrow \bar\psi\,.
\eeq


\section{Double transformations of the fields}
\label{app:conventionsanddoubletransformations}

The main ingredient in the construction of fermionic Wilson loops are the double transformations of scalar fields. Using \eqref{SUSY2}, we write these as
\begin{align}
\label{double}
[\delta_1,\delta_2]q_I^a
&=-i(\xi_1^{a\dot b}\gamma^\mu\xi_{2b\dot b}-\xi_2^{a\dot b}\gamma^\mu\xi_{1b\dot b}) D_\mu q_I^b
-i(\xi_1^{a\dot b}\zeta_{2b\dot b}-\xi_2^{a\dot b}\zeta_{1b\dot b})q_I^b
\nonumber
\\&\quad
+\frac{i}{k}(\xi_1^{a\dot b}\xi_{2b\dot b}-\xi_2^{a\dot b}\xi_{1b\dot b})(\nu_Iq_I^b-q_I^b\nu_{I+1})
-\frac{2i}{k}(\xi_1^{a\dot a}\xi_{2b\dot b}-\xi_2^{a\dot a}\xi_{1b\dot b})
(\tilde\mu_I{}^{\dot b}_{\ \dot a}q_I^b-q_I^b\tilde\mu_{I+1}{}^{\ \dot b}_{\dot a})
\nonumber
\\
[\delta_1,\delta_2]\tilde q_{I-1\,\dot a}
&=-i(\xi_{1 a\dot a}\gamma^\mu\xi_{2}^{a\dot b}-\xi_{2 a\dot a}\gamma^\mu\xi_{1}^{a\dot b}) D_\mu \tilde q_{I-1 \dot b}
-i(\xi_{1 a\dot a}\zeta_{2}^{a\dot b}-\xi_{2 a\dot a}\zeta_{1}^{a\dot b})\tilde q_{I-1 \dot b}
\cr&\quad
+\frac{i}{k}(\xi_{1 a\dot a}\xi_{2}^{a\dot b}-\xi_{2 a\dot a}\xi_{1}^{a\dot b})
(\tilde q_{I-1 \dot b}\tilde\nu_I-  \tilde\nu_{I-1}\tilde q_{I-1 \dot b})
\nonumber
\\&\quad
-\frac{2i}{k}(\xi_{1 a\dot a}\xi^{b\dot b}_2-\xi_{2 a\dot a}\xi^{b\dot b}_1)
(\tilde q_{I-1 \dot b}\mu_I{}^{a}_{\ b}-\mu_{I-1}{}^{a}_{\ b} \tilde q_{I-1 \dot b})\,.
\end{align}

We specialize the double transformations to the supercharges (\ref{supercharges}), whose corresponding parameters are given by
\bea
\xi_{1,a\dot b}=\delta_a^1\left(\delta_{\dot b}^{\dot 1}\cos\frac{\theta}{2}\xi^{\bar l}+\delta_{\dot b}^{\dot 2}\sin\frac{\theta}{2}\xi^{\bar r}\right)\,, \qquad
\xi_{2,a\dot b}=\delta_a^2\left(\delta_{\dot b}^{\dot 2}\cos\frac{\theta}{2}\xi^l-\delta_{\dot b}^{\dot 1}\sin\frac{\theta}{2}\xi^r\right)\,.
\eea
The Killing spinors in (\ref{killingspinors}) obey (\ref{xi}), from which one sees that the second term in (\ref{double}) becomes
\bea
-i(\xi_1^{a\dot b}\zeta_{2b\dot b}-\xi_2^{a\dot b}\zeta_{1b\dot b})q_I^b &=&
-\frac{1}{2}\cos\theta \left(\delta^a_1 q^1_I-\delta^a_2  q^2_I\right)
\,.
\eea
Combining with the rest of (\ref{double}) one finds
\bal
i\{ \cQ_ 1,\cQ_2\} q^1_I&=D_\varphi q^1_I -\frac{i}{2}\cos\theta\, q^1_I
+\frac{1}{k}(\nu_I q^1_I-q_I^1 \nu_{I+1})
\\
&\quad{}-\frac{1}{k}\left(\bar{\tilde r}^{\,\dot 1}_{I-1}\tilde{r}_{I-1\,\dot 1}-\bar{\tilde r}^{\,\dot 2}_{I-1}\tilde{r}_{I-1\,\dot 2}\right)q^1_I
+\frac{1}{k}q^1_{I}\left(\tilde{r}_{I+1\,\dot 1}\bar{\tilde r}^{\,\dot 1}_{I+1}-\tilde{r}_{I+1\,\dot 2}\bar{\tilde r}^{\,\dot 2}_{I+1}\right)\,,
\\
i\{ \cQ_ 1,\cQ_2\} q^2_I&= D_\varphi q^2_I +\frac{i}{2}\cos\theta\, q^2_I
-\frac{1}{k}(\nu_I q^2_I-q_I^2 \nu_{I+1})
\\
&\quad{}-\frac{1}{k}\left(\bar{\tilde r}^{\,\dot 1}_{I-1}\tilde{r}_{I-1\,\dot 1}-\bar{\tilde r}^{\,\dot 2}_{I-1}\tilde{r}_{I-1\,\dot 2}\right)q^2_I
+\frac{1}{k}q^2_{I}\left(\tilde{r}_{I+1\,\dot 1}\bar{\tilde r}^{\,\dot 1}_{I+1}-\tilde{r}_{I+1\,\dot 2}\bar{\tilde r}^{\,\dot 2}_{I+1}\right)
\,,
\eal
with fields $\tilde r$ defined in \eqref{rotatedscalarfields}.
Noting that
\bal
\nu_I q_I^1-q_I^1\nu_{I+1}&
=-(\mu_I{}^1_{\ 1}-\mu_I{}^2_{\ 2})q_I^1+q_I^1(\mu_{I+1}{}^{\ 1}_{1}-\mu_{I+1}{}^{\ 2}_{2})\,,
\\
\nu_I q_I^2-q_I^2\nu_{I+1}
&
=(\mu_I{}^1_{\ 1}-\mu_I{}^2_{\ 2})q_I^2-q_I^2(\mu_{I+1}{}^{\ 1}_{1}-\mu_{I+1}{}^{\ 2}_{2})
\eal
we can write
\beq
i\{ \cQ_ 1,\cQ_2\} q^a_I=D_\varphi q^a_I -\frac{i}{2}\cos\theta \left(\delta^a_1 q^1_I-\delta^a_2 q^2_I\right)
-i(\cA_I^\theta- A_{\varphi,I})q^a_I+iq^a_I(\cA_{I+1}^\theta- A_{\varphi,I+1})
\,. 
\eeq
where $\cA_{I+1}^\theta$ is given by the natural generalization of \eqref{rotated-cA} to the $I+1$-th node
\beq
\label{AIpm1}
\cA^{\theta}_{I\pm1}=A_{\varphi,I\pm1}
-\frac{i}{k}({{\mu_{I\pm1}}_1}^{1}-{{\mu_{I\pm1}}_2}^{2}
+\tilde r_{I\pm1\,\dot1}\bar{\tilde r}_{I\pm1}^{\,\dot 1}-\tilde r_{I\pm1\,\dot2}\bar{\tilde r}_{I\pm1}^{\,\dot2})\,.
\eeq
The covariant derivative is $D_\varphi q^a_I = \partial_\varphi q^a_I-i A_{\varphi,I}q^a_I+iq^a_I A_{\varphi,I+1}$, so that the double transformation can be recast as a total covariant derivative with respect to the ${\cal A}^\theta$-connection, including the coupling to a background field
\bal
\label{Q1Q2q}
i\{ \cQ_ 1,\cQ_2\} q^1_I&=\cD_\varphi q^1_I\equiv \partial_\varphi q^1_I -\frac{i}{2}\cos\theta q^1_I
-i\cA_I^\theta q^1_I+iq^1_I \cA_{I+1}^\theta \,,
\\
i\{ \cQ_ 1,\cQ_2\} q^2_I&=\cD_\varphi q^2_I\equiv \partial_\varphi q^2_I +\frac{i}{2}\cos\theta q^2_I
-i\cA_I^\theta q^2_I+iq^2_I \cA_{I+1}^\theta \,.
\eal
Note that $q^1_I$ and $q^2_I$ are charged oppositely with respect to the background field.

We repeat the computation for the $\tilde r_{\dot a}$'s arriving at
\bal
i\{ \cQ_ 1,\cQ_2\} \tilde r_{I-1\,\dot 1}&=D_\varphi \tilde r_{I-1\,\dot 1}+\frac{i}{2}\tilde r_{I-1\,\dot 1}
- \frac{1}{k}(\tilde r_{I-1\,\dot 1}\tilde \nu_I - \tilde\nu_{I-1}\tilde r_{I-1\,\dot 1})\\
&\quad{}+\frac{1}{k}\left(\tilde r_{I-1\,\dot 1}({{\mu_{I}}^1}_1-{{\mu_{I}}^2}_2) -({\mu_{I-1\,  1}}^1-{\mu_{I-1\,  2}}^2) \tilde r_{I-1\,\dot 1} \right)\,,\\
i\{ \cQ_ 1,\cQ_2\} \tilde r_{I-1\,\dot 2}&=  D_\varphi \tilde r_{I-1\,\dot 2}-\frac{i}{2}\tilde r_{I-1\,\dot 2}
+\frac{1}{k}(\tilde r_{I-1\,\dot 2}\tilde \nu_I - \tilde\nu_{I-1}\tilde r_{I-1\,\dot 2})\\
&\quad{}+\frac{1}{k}\left(\tilde r_{I-1\,\dot 2}({{\mu_{I}}^1}_1-{{\mu_{I}}^2}_2) -({\mu_{I-1\,  1}}^1-{\mu_{I-1\,  2}}^2)  \tilde r_{I-1\,\dot 2} \right)\,.
\eal
It is easy to see that these are also given by total derivatives
\bal
\label{Q1Q2tildet}
i\{ \cQ_ 1,\cQ_2\} \tilde r_{I-1\, \dot 1}&=
\cD_\varphi  \tilde r_{I-1\, \dot 1} \equiv \partial_\varphi \tilde r_{I-1\, \dot 1} +\frac{i}{2}\tilde r_{I-1\, \dot 1} 
-i \cA_{I-1}^\theta\tilde r_{I-1\, \dot 1}+i\tilde r_{I-1\, \dot 1} \cA_I^\theta  \,,
\\
i\{ \cQ_ 1,\cQ_2\} \tilde r_{I-1\, \dot 2}&=
\cD_\varphi  \tilde r_{I-1\, \dot 2} \equiv \partial_\varphi \tilde r_{I-1\, \dot 2} -\frac{i}{2}\tilde r_{I-1\, \dot 2}
-i \cA_{I-1}^\theta\tilde r_{I-1\, \dot 2}+i\tilde r_{I-1\, \dot 2} \cA_I^\theta  \,,
\eal
with, again, the two components oppositely charged with respect to the background field. We also need the corresponding expressions for the conjugate fields, which are given by
\bal
i\{\cQ_1,\cQ_2\}\bar{q}_{I a}&=\cD_\varphi \bar{q}_{I a}\equiv \partial_\varphi \bar{q}_{I a} 
-(-1)^a\frac{i}{2}\cos\theta\,\bar{q}_{I a}-i\cA_{I+1}^\theta\bar{q}_{I a}+i\bar{q}_{I a}\cA_I^\theta
\,,\\
i\{\cQ_1,\cQ_2\}\bar{\tilde{r}}_{I-1}^{\,\dot{a}}&=\cD_\varphi \bar{\tilde{r}}_{I-1}^{\,\dot{a}}\equiv \partial_\varphi \bar{\tilde{r}}_{I-1}^{\,\dot{a}} 
+(-1)^{\dot a}\frac{i}{2}\bar{\tilde{r}}_{I-1}^{\,\dot{a}}
-i\cA_{I}^\theta\bar{\tilde{r}}_{I-1}^{\,\dot{a}}+i\bar{\tilde{r}}_{I-1}^{\,\dot{a}}\cA_{I-1}^\theta\,. \label{Q1Q2bartildet}
\eal


\bibliographystyle{utphys2}
\bibliography{refs}
\end{document}